# Understanding the Role of Non-Fullerene Acceptors Crystallinity on the Charge Transport Properties and Performance of Organic Solar Cells

Pierluigi Mondelli,[a,d,†,*] Pascal Kaienburg,[a] Francesco Silvestri,[b] Rebecca Scatena,[a] Claire Welton,[c] Martine Grandjean,[d] Vincent Lemaur,[e] Eduardo Solano,[f] Mathias Nyman,[g] Peter N. Horton,[h] Simon J. Coles,[h] Esther Barrena,[b] Moritz Riede,[a] Paolo Radaelli,[a] David Beljonne,[e] G. N. Manjunatha Reddy[c] and Graham Morse[d,†]

The acceptor crystallinity has long been associated with favourable organic solar cells (OSCs) properties such as high mobility and Fill Factor. In particular, this applies to acceptor materials such as fullerene-derivatives and the most recent Non-Fullerene Acceptors (NFAs), which are now surpassing 19% of Power Conversion Efficiency. Despite these advantages are commonly attributed to their 3-dimensional crystal packing motif in the single crystal, the bridge that links the acceptor crystal packing from single crystals to solar cells has not clearly been shown yet. In this work, we investigate the molecular organisation of seven NFAs (o-IDTBR, IDIC, ITIC, m-ITIC, 4TIC, 4TICO, m-4TICO), following the evolution of their packing motif in single-crystals, powder, and thin films made with pure NFAs and donor:NFA blends. We observed a good correlation between the NFA single crystal packing motif and their molecular arrangement in the bulk heterojunction. The NFA packing motif affects the material's propensity to form highly crystalline domain in the blend. We specifically found that 3D reticular packing motifs show stronger ordering than 0D herringbone ones. However, the NFA packing motif is not directly correlating with device performance parameters: Although higher NFA crystallinity yields higher mobility, we found the domain purity to be more important for obtaining high efficiency organic solar cells by governing bimolecular recombination.

## Introduction

The recent surge in Organic Solar Cells (OSCs) performance, now exceeding 19%,[1, 2] results from the development of Non-Fullerene Acceptors (NFAs).[3-8] Previous work centred around fullerene based acceptors has drawn a connection between molecular design/shape to the formation of highly-interconnected acceptor domains and charge percolation pathways towards the electrodes, resulting in superior charge transport properties and high Fill Factors (FF) in OSC.[9-12] Moreover, recent works are attributing the improved performance and charge transport of state of the art NFAs to their 3D-interconnected crystal packing motif (Figure 1).[5, 7-9, 13-17] It is thus important to construct a clear understanding based on concrete evidence of the relationship between the NFA molecular packing and crystallinity in the bulk heterojunction and the charge transport properties and performance of organic solar cells.

A growing variety of NFA single-crystals is now being reported,[8, 14, 15, 18-31] from which useful information on the molecular packing can be derived. Still, bridging the molecular scale from molecular packing in single-crystals to the one found in Bulk Heterojuction (BHJ) needs more detailed investigations.[10, 31, 32] The identification of the packing motif of the NFA within a donor:acceptor blend is a challanging task, yet the structural analysis is commonly based on Grazing Incidence Wide Angle X-Ray Scattering (GIWAXS) patterns that lack of enough diffraction features for the structural determination. The analysis of the main Bragg peaks of the NFA, often referred as the lamellar (100) and the π-π stacking (010) distances,[33-35] is routinely considered enough to draw conclusions. Misleading results can, however, arise from the similar (if not overlapping) spectral features of the donor and acceptor components in the q-space, the typical peak broadening of organic compounds reflecting a high degree of flexibility, the presence of rotational isomers, polymorphism,[36, 37] and a general lack of long-range crystalline order.[34]

In this work, we used an extensive set of both experimental and theoretical approaches to study the molecular packing and morphology of a specific set of common NFAs (o-IDTBR, IDIC, ITIC, m-ITIC, 4TIC, 4TICO, m-4TICO, whose structures are shown in Figure 2), by means of single-crystal X-Ray Diffraction (XRD), powder XRD and Le Bail refinement, GIWAXS, Atomic Force Microscopy (AFM), crystal lattice simulations, ss-NMR, and Gauge Including Projected Augmented Wave (GIPAW) DFT calculations. Our aim was to analyse the NFA molecular packing,

[a.] Clarendon Laboratory, Department of Physics, University of Oxford, Parks Road, Oxford, OX1 3PU, United Kingdom. E-mail: pierluigi.mondelli@physics.ox.ac.uk
[b.] Institut de Ciència de Materials de Barcelona, ICMAB-CSIC, Campus UAB, 08193 Bellaterra, Spain
[c.] University of Lille, CNRS, Centrale Lille, Univ. Artois, UMR 8181- UCCS - Unité de Catalyse et Chimie du Solide, F-59000 Lille, France
[d.] Merck Chemicals Ltd, Chilworth Technical Centre, University Parkway, Southampton, SO16 7QD, United Kingdom
[e.] Laboratory for Chemistry of Novel Materials, University of Mons, Place du Parc, 20, 7000 Mons (Belgium)
[f.] NCD-SWEET beamline, ALBA Synchrotron Light Source, Cerdanyola del Vallès, 08290 Spain
[g.] Physics, Faculty of Science and Engineering, Åbo Akademi University, 20500 Turku, Finland
[h.] EPSRC Crystallographic Service, Department of Chemistry, University of Southampton, Highfield, SO17 1BJ, UK
*Current address: Italian Institute of Technology, Printed and Molecular Electronics, Via Pascoli 70/3, 20133 Milan, Italy
† Email: pierluigi.mondelli@gmail.com, graham.morse@gmail.com

## Molecular Packing

0D - Herringbone    2D - Brickwork    3D - Reticular

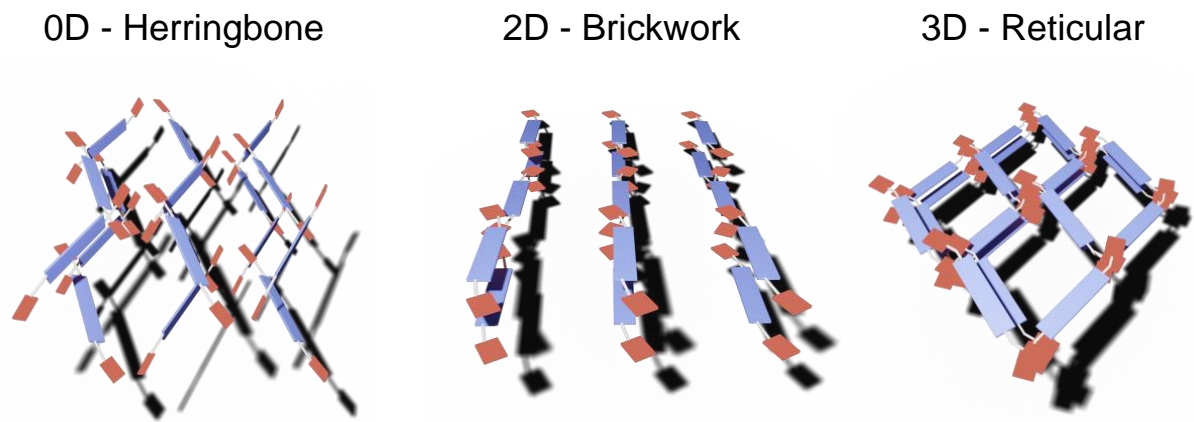

Figure 1. Sketch of the different molecular packing motifs observed in Acceptor-Donor-Acceptor (A-D-A) type NFAs crystal structures and labelled according to the dimensionality of the π-π stacking.

with a special focus on the structural evolution from single crystals to solar cells. We studied the molecular arrangement from single crystals, to powders, then thin films and finally morphology in the bulk heterojunction (BHJ) to bring important insights into the NFA arrangement in the BHJs. We analysed the most relevant NFA packing motifs and polymorphs, discussing their role in the solar cell morphology, current-voltage performance, and charge transport properties, such as electron mobility and bimolecular recombination, by means of photo-CELIV (Charge Extraction Linear Increasing Voltage) and MIS-CELIV (Metal-Insulator-Semiconductor-Charge Extraction Linear Increasing Voltage).

## Experimental

### General Characterisation

UV-Vis absorption spectra were recorded on an OceanOptics QE PRO spectrometer using a tungsten halogen light source (HL-2000-FHSA from OceanOptics). The HOMO levels of the compounds in thin films were obtained by measuring the film photoemission current onset with the Air Photoemission Spectroscopy module (APS02) of the Kelvin Probe from KP Technology Ltd. The Kelvin Probe was equipped with a 2.0 mm diameter tip coated with gold alloy, a UV deuterium lamp, and a monochromator (range 3.44-3.88 eV). LUMO levels were estimated by adding the optical bandgap (determined from the onset of the UV-Vis absorption) to the measured HOMO.

Single-crystal X-ray diffraction data were collected on a Rigaku Oxford Diffraction Supernova diffractometer equipped with micro-focus sealed (Mo-Kα) X-ray Source, CCD plate detector and Oxford Cryostream N2 flow cryostat. The samples were mounted on Kapton loops from the solution and shock-cooled to 173.0(2) K or 100.0(2) K. Cell indexing and peak integration were performed with CrysAlisPro. Structural solution and refinement were carried out with ShelxT and ShelxL, respectively.

Powder X-ray diffraction was performed using a PANalytical X'Pert PRO diffractometer with Cu -Kα radiation. During the measurement, the sample was kept at room temperature and under ambient conditions.

### GIWAXS

The GIWAXS experiments were carried out at NCD-SWEET beamline at ALBA synchrotron (Beamtime ID: 2019093873). A monochromatic X-ray beam with a photon energy of 12.4 keV was set using a Si (1 1 1) channel cut monochromator, further collimated with an array of beryllium lenses. The GIWAXS maps were recorded with a Rayonix LX 255-HS detector, consisting of a pixel array of 960 × 2880 pixels of 88.54 × 88.54 μm$^2$ (H × V) for the binning employed. The samples were thermally annealed before the measurements at the optimised temperature for the solar cell performance (see below) using a hotplate in air. GIWAXS frames were acquired near the critical angle of the glass substrate (ca. 0.15° for the X-ray wavelength employed), penetrating a depth of 11 nm for the layer of interest[38] while minimizing the contribution of the substrate.[‡] The recorded 2D scattering patterns were analysed using a home-made python routine based on pyFAI (the Fast Azimuthal Integration Python library).[39] GIWAXS images are in logarithmic scale, ranging from dark blue (low intensity) to yellow (high intensity). The in-plane and out-of-plane profiles were obtained by integrating the diffraction intensity in rectangular areas centred at $\chi=0°$ and $\chi=90°$ from $\chi$-$q$ images (scattering intensity as a function of the azimuthal angle). The scattering peaks of the bulk structure were compared to the experimental data using SimDiffraction, a MatLab code for simulating the film diffraction pattern for a given crystal structure and orientation.[40] For each NFA, The simulations were performed by choosing a specific NFA orientation with respect to the substrate (typically in-plane and out-of-plane). The Miller indices ($h$ $k$ $l$) associated with the NFA packing direction were determined by Mercury[41] and used as input parameters for the simulations. A better fit between simulated and experimental GIWAXS was obtained when using the unit cell parameters obtained by Le Bail refinement (see below) as input for the simulations.

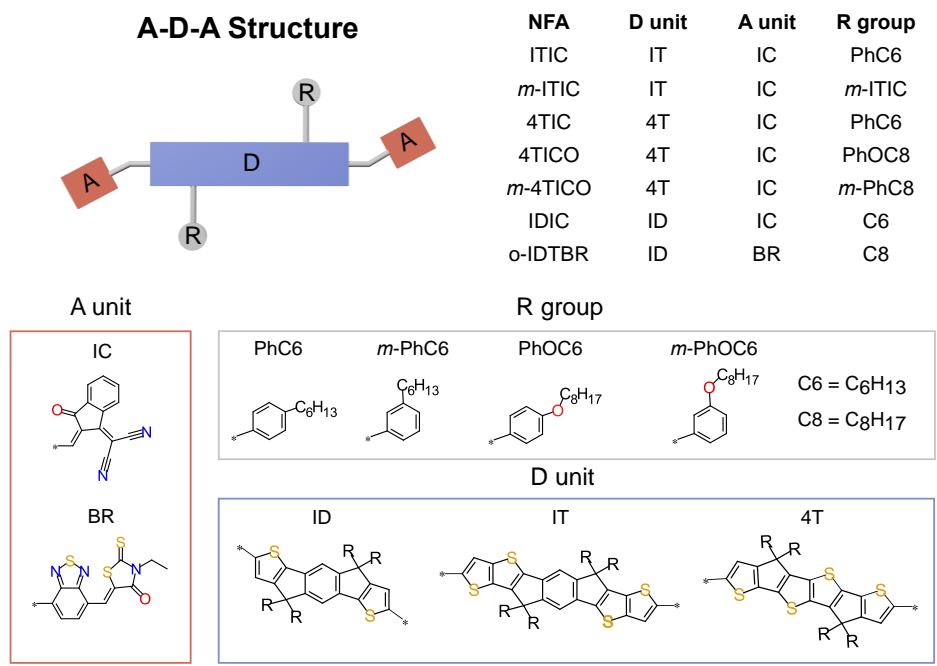

Figure 2. Structure of an Acceptor-Donor-Acceptor (A-D-A) NFA with its building blocks on the top left. The NFAs studied in this work (table on the top right) are identified with their chemical subunits (A, D and R groups) whose chemical structures are also shown.

**Solid-state NMR spectroscopy**

For ssNMR experiments, O-IDTBR powder was packed into a 1.3 mm (outer diameter) rotor. All 1D $^1$H and $^{13}$C[20], and 2D $^1$H-$^1$H and $^1$H-$^{13}$C correlation NMR experiments were carried out on a Bruker Avance Neo (18.8 T) spectrometer using a 1.3 mm double resonance H-X probehead tuned to $^1$H (Larmor frequency, 800.13 MHz) and $^{13}$C (Larmor frequency, 201.2 MHz) nuclei. Unless otherwise states, the Magic Angle Spinning (MAS) frequency was 50 kHz in all cases. The nutation frequencies for $^1$H and $^{13}$C were 100 kHz and 90 kHz, corresponding to 90º pulse durations of 2.5 and 2.75 μs, respectively. The longitudinal relaxation time ($T_1$) of $^1$H was determined to be 3 s based on inversion recovery measurements and analyses. 1D $^1$H MAS NMR spectrum was acquired using 16 co-added transients. A 2D $^1$H-$^1$H double-quantum (DQ)-single-quantum (SQ) NMR spectrum was acquired using Back-to-Back (BaBa) sequence at fast MAS,[42] using a rotor-synchronized $t_1$ increment of 20 μs corresponding to one rotor period ($\tau_r$). The indirect $^1$H DQ dimension was acquired using 256 $t_1$ increments, each with 16 co-added transients, corresponding to a total experimental time of ~4 h. $^1$H detected 2D $^1$H-$^{13}$C heteronuclear correlation (HETCOR) spectra were acquired with 0.1 ms and 3 ms of CP contact time and the indirect $^{13}$C dimension was acquired using 140 $t_1$ increments, each with 32 co-added transients, corresponding to a total experimental time of 8 h each.

**Atomic Force Microscopy**

The thin films were investigated by AFM both in contact and dynamic modes using a commercial head and control unit from Nanotec Electronica. The used thermal annealing protocol was the one optimized for the devices (see Table S10). For each sample, after each annealing step, different spots of the surface were imaged (at several image sizes) to have a statistical validity of the measurements. The images presented in the article are chosen as high-resolution representative images of the surface. The estimation of the root mean square (RMS) roughness was done selecting about 6 contact mode images (30x30 and 50x50 μm²) for each temperature. Si$_3$N$_4$ V-shaped cantilevers (Veeco) with the nominal force constant ranging between 0.03 and 0.5 N/m were employed for the contact mode, while Cr/Pt-coated silicon tips on rectangular cantilevers (BudgetSensors) with a nominal resonance frequency of 75 kHz and a force constant of 3 N/m were used for the dynamic mode. The open-source Gwyddion software was used to analyse all the presented AFM images,[43] including the domain size calculation which were done using the watershed algorithm.[44] The input parameters used for the watershed analysis can be found in the Supporting Information (pages S20-S25).

**Solar Cells Fabrication and Characterisation**

Inverted-architecture organic solar cells were fabricated by blade coating of the organic layers on Indium Tin Oxide(ITO)/glass pre-patterned 5x5 cm² substrates (Zencatec Limited). A detailed description of the interlayers and electrodes fabrication, along with the experimental description of the current-voltage (I-V) measurements can be found in the experimental section of a recent work from our group.[36] Aluminium-doped zinc oxide from Avantama (N-21X-Slot) was used for the electron transporting layer, while PEDOT:PSS (Clevios Al 4083 from Heraeus) for the hole transporting layer. All the NFAs were supplied by 1-Material Inc., with the only exception of 4TICO (Merck KGaA). For the active layer, PBTZT-stat-BDTT-8 (Merck KGaA) was used as the donor material[45] in combination with the NFAs listed in Figure 2. Each blend was dissolved in a 1:1.3 ratio (by weight) and 80 nm thick layers[‡‡] were processed from a 23 mg/ml solid content o-xylene solution, without the use of additional additives. The blade speed was adjusted between 7 to 13 mm s$^{-1}$ to reach the desired thickness with a 100 μm blade gap and 70 μL cast volume. Casting plate temperature was varied between 60°C and 80°C and the as-cast devices were further annealed at temperatures ranging from 100°C to 140°C on a hot plate in air following the optimisation protocol. A 100 nm silver back electrode was thermally evaporated on top of the hole transporting layer, under a pressure of 2x10$^{-6}$ mbar at a rate of 1 Å/s. Solar cells have a device area of ~ 8.5 mm$^2$, as determined from the geometrical overlap between cathode and anode.

**Photo-CELIV**

Photo-CELIV was performed using the Transient Measurement Unit by Automatic Research. A 655 nm laser pulse (5 μs long) was used to photo-generate charges into ~ 8.5 mm$^2$ area solar cells. The linear increasing voltage ramp range was 0 - 1.5 V and 10 μs long. The delay time between the laser pulse and the voltage ramp was varied between 0.1 to 100 μs, during which a constant pre-bias voltage (close to $V_{OC}$) was applied to limit the charge injection. The charge mobility was determined by using eq.(1) of ref.[46] on the 30 μs long delay time data collection. The bimolecular recombination coefficient was calculated using eq.(2) of ref.[47]

**MIS-CELIV**

MIS-CELIV[48] was performed with a PAIOS system from Fluxim AG, Switzerland. The layer stack was ITO/AZO/active layer/MgF$_2$/Ag with an insulating 50 nm MgF$_2$ layer that blocks injection of holes so that the electron mobility is obtained. The chosen thickness balances the layer's insulating properties with its capacitance in relation to the absorber's capacitance. The other layers are processed as for the solar cell devices with some deviations in the annealing protocol discussed in the Support Information. A small device area of ~1mm$^2$ was chosen to minimize RC effects. The offset voltage was typically varied between 0V and 8V and the ramp rate between 50 V/ms and 1600 V/ms. The latter allowed to account for injection barrier effects[49]. The analysis was carried out following the diffusion-corrected eq.(11) of ref.[49] with saturation and geometric displacement current densities extracted from the CELIV curves.

**Density Functional Theory calculations**

Periodic DFT calculations on the o-IDTBR crystalline structure have been performed using the CASTEP module with the Materials Studio software. All calculations have been carried out with the PBE GGA functional, a plane-wave energy cutoff of 50 Rydbergs (680 eV) and a k-point spacing of 0.05Å$^{-1}$.[50] The crystalline structure of o-IDTBR has first been full relaxed using the Tkatchenko-Scheffler dispersion correction method, optimizing both all atomic positions within the cell and unit cell parameters. The resulting DFT-optimized cell parameters (a = 13.8706Å, b = 15.5913Å, c = 32.6925Å, α = 90°, β = 96.0529, γ = 90°) are in excellent agreement with the measured crystallographic data in Table 1. NMR calculations have then been performed on the optimized crystal structure using the Gauge-Including Projector Augmented-Wave method (GIPAW); reference shieldings of 31.09 and 179.02 ppm were used for $^1$H and $^{13}$C, respectively.

## Results and Discussion

### Structural Analysis

A common way of classifying the organic semiconductor packing motif is by observing their π-π stacking dimensionality.[18, 51-54] A-D-A molecules (Figure 2) in particular, can form highly interconnected domains through intermolecular interactions between the acceptor units (A units) of adjacent molecules.[18, 55] The percolation pathway that forms through π-π stacking can develop along multiple directions of the crystalline domain. Molecules can arrange through brickwork pattern with 2D percolation pathways, or through the so-called "reticular" packing motif which is characterised by 3D-interconnected domains. For herringbone crystal structures the molecular backbones of adjacent units are orthogonal, therefore lack π-π stacking (Figure 1).

To reveal the NFA packing within the BHJ we started our investigations from single crystals, which represent the perfect platform to explore the influence of the solid-state arrangement on the charge transport. Most of the crystal structures analysed in this work were previously resolved by our group,[18] while others were found in literature. As indicated in Table 1, some NFAs showed polymorphism. For instance, ITIC single crystals can be found to be either 0D herringbone or 2D brickwork motifs and *m*-4TICO also presents two different unit cells.

Single crystals represent the NFA molecular ordering of high purity and large millimetre sized crystal grown from controlled conditions (solvent vapour diffusion, see reference[18]) and measured by XRD in a low-temperature (100 K) environment. Thus, it is possible to observe substantial differences in the molecular arrangement as we deviate from such ideal systems towards the BHJ. Therefore, we wanted to understand how the crystal packing changes by raising the temperature to ambient conditions (200 K shift between single crystal and powder XRD) and by disrupting the ideal growth conditions and long-range crystallinity of NFA single crystals. As an intermediate step, ss-NMR and XRD of purified NFA powders were performed. In the main text, we limit our observations on *o*-IDTBR, extending the analysis and discussion to the other materials in the Supplementary Information (pages S6-S9).

Table 1. Crystallographic information of the NFA crystal structures available for the materials analysed in this work.

| CCDC Identifier | Molecule | Motif | π-π | a (Å) | b (Å) | c (Å) | α (deg) | β (deg) | γ (deg) | Volume (Å³) |
|---|---|---|---|---|---|---|---|---|---|---|
| FOSPOV[56] | o-IDTBR | reticular | 3D | 13.7663(2) | 15.8103(17) | 32.7146(3) | 90 | 96.2928(12) | 90 | 7077.43(15) |
| YEBKEY[19] | 4TIC | reticular | 3D | 13.969(7) | 17.144(9) | 17.970(10) | 104.668(16) | 109.998(17) | 96.169(14) | 3822.08 |
| VUBJIO[18] | m-4TICO | brickwork | 2D | 8.6526(3) | 16.4878(8) | 18.0435(8) | 114.697(5) | 103.822(4) | 90.890(4) | 2251.45(19) |
| This work | m-4TICO | brickwork | 2D | 8.7845(7) | 15.3726(13) | 16.7896(13) | 67.136(7) | 85.678(7) | 79.630(7) | 2054.98 |
| VUBJOU[18] | m-ITIC | brickwork | 2D | 8.7454(13) | 18.872(2) | 25.2647(18) | 87.770(8) | 88.724(9) | 78.001(12) | 4075.1(9) |
| VUBKAH[18] | IDIC | brickwork | 2D | 8.6679(4) | 12.5073(7) | 13.5784(6) | 72.096(4) | 75.545(4) | 88.839(4) | 1353.88(12) |
| VUBJEK[18] | 4TICO | herringbone | 0D | 15.2836(2) | 20.0101(5) | 29.3242(6) | 90 | 89.997(2) | 90 | 8968.1(3) |
| KIZSUK[20] | ITIC | brickwork | 2D | 8.420(6) | 23.019(17) | 23.126(17) | 101.780(10) | 95.319(10) | 91.105(14) | 4366(5) |
| HEHQUJ01[18] | ITIC | herringbone | 0D | 14.9009(7) | 15.5043(4) | 18.1199(5) | 99.309(2) | 101.541(3) | 108.366(3) | 3777.2(2) |

**From Single Crystals to Powder**

o-IDTBR has a 3D reticular packing motif in the single crystal,[56] characterised by close-contacts between the electron accepting units (A units) of tilted adjacent molecules (Figure 1). To understand if this packing geometry is retained in powder samples, we performed X-Ray Diffraction and ss-NMR measurements. The powder diffractogram showed a long-range crystallinity with well-defined Bragg peaks in the low-angle region (Figure 3a). A partial agreement exists between the experimental data and the simulated 1D pattern of the single crystal structure, yet a certain mismatch between the positions for most of the reflections suggests that a slightly different unit cell is formed in powder. This can occur because of the

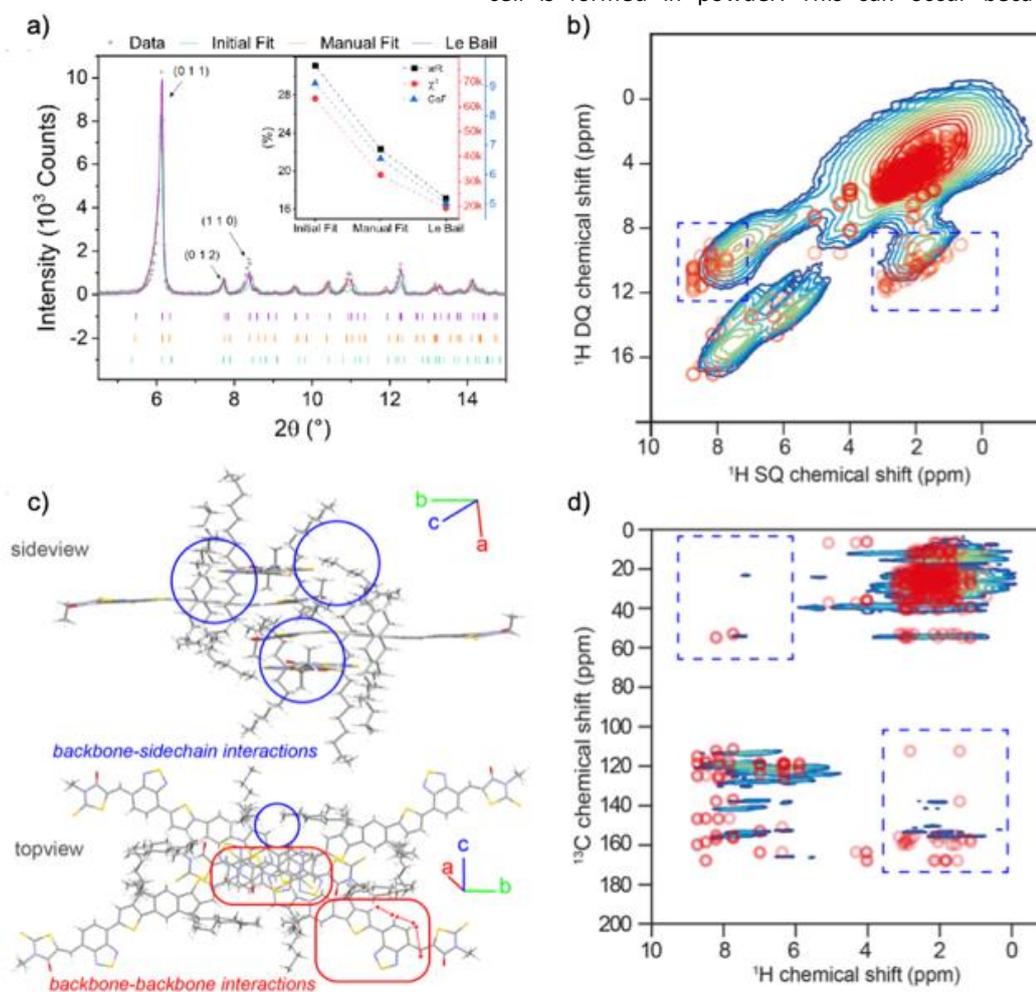

Figure 3 a) Powder XRD diffractogram overlaid to the simulated powder pattern of the single crystal structure. The inset is showing the progressive improvement of the fitting from the original fit to manual and Le Bail refinement. c) Periodic DFT optimised crystal structure and highlighted backbone-backbone (red boxes) and backbone-sidechains interaction (blue boxes). b-d) DFT calculated 2D plots of the $^1$H-$^1$H and $^1$H-$^{13}$C chemical shifts (red circles) overlaid on the experimental $^1$H-$^1$H double-quantum-single-quantum (DQ-SQ) correlation and $^1$H-$^{13}$C heteronuclear correlation spectra (contours), respectively. The dashed blue rectangles indicate the minor changes in the backbone-sidechain interactions when comparing the crystalline to powder form, meaning that the π-π interactions (along with the packing motif) remains substantially unvaried.

temperature shift (200 K) between the single crystal and powder XRD measurements,§ which can impact the long-range order, the local structures and packing interactions. In a first approximation,§§ we manually solved the reciprocal-space metric tensor equation for monoclinic structures[57] to derive the lattice parameters from the experimental data (Table S1). The spectral agreement between the "manually-refined" simulated diffractogram and the experimental data improves (Figure 3a), as the Goodness of Fit (GOF), Chi$^2$ and residuals (wR) parameters are now reduced. This means that a better fit is obtained after the manual refinement, and therefore it is reasonable to assume a structural agreement between the two phases. A further confirmation is obtained when the lattice parameters and unit cell angles are derived through Le Bail refinement, suggesting that the structure undergoes a volumetric expansion from single crystal to powder phase (Table S1), possibly resulting from the different temperatures of the measurements (see Supporting Information, Figure S5 and Table S2) causing subtle changes in the local interactions.

To further confirm this behaviour, we performed NMR crystallography analysis, which combines XRD, ssNMR spectroscopy and modelling (here first principles calculations and GIPAW-DFT based NMR chemical shielding calculations) techniques to resolve atomic-scale interactions.[58] Solid-state NMR is particularly sensitive to local structures of polymeric organic semiconductors, NFAs and polymer:NFA blends.[59-61] Here, we carried out ssNMR crystallography analysis with the aim of identifying the changes in local structures in crystals and powder compositions (Figure 3b-d). This is achieved by analysing and comparing $^1$H and $^{13}$C chemical shifts of crystal structures as calculated by GIPAW-DFT approach with the experimentally measured $^{13}$C and $^1$H chemical shifts for the o-IDTBR powder. A detailed analysis of experimental 1D $^1$H, $^{13}$C and 2D $^1$H-$^1$H and $^1$H-$^{13}$C correlation ssNMR spectra is presented in Supporting Information (Figures S1-S4). Periodic DFT optimised crystal structures are shown in Figure 3c, whereby the backbone-backbone and backbone-sidechain interactions are indicated in soft-rectangles (in red) and circles (in blue), respectively. Figure 3b,d compares the 2D plots of DFT-calculated chemical shifts generated by MagresView and MagresPython software tool[62] for $^1$H-$^1$H and $^1$H-$^{13}$C spin pairs within a 3 Å distance, overlaid on the experimental $^1$H-$^1$H double-quantum-single-quantum (DQ-SQ) correlation and $^1$H-$^{13}$C nuclear correlation spectra. In 2D NMR measurements of this type, 2D peaks corresponding to $^1$H-$^1$H and $^1$H-$^{13}$C proximities within sub-nanometre distances in powder solids are detected. It is noteworthy that a good correlation between the GIPAW-DFT calculated chemical shifts and the experimental chemical shifts ($\delta$) are observed for both aliphatic and aromatic moieties. In the DQ-SQ spectrum (Figure 3b, the broad DQ peak at 0-8 ppm in the vertical axis is due to the $^1$H-$^1$H proximities in alkyl sidechains and the DQ peaks in 12-16 ppm range are due to the through-space $^1$H-$^1$H proximities between aromatic groups within the chain and in between the $\pi-\pi$ stacked o-IDTBR molecules, both of which exhibit good agreement with the DFT-calculated chemical shifts. However, subtle differences between the DFT calculated and experimental chemical shifts $\delta(^1H_{DQ})$ in the 8-12 ppm range (dashed blue boxes), which originate from through-space dipolar interactions between aromatic groups and sidechains, indicate the minor changes in the backbone-sidechain interactions in the vicinity of CH$_2$ moieties when compared the crystalline and powder forms.[63] Similarly, a good agreement is obtained when comparing the DFT-calculated chemical shifts of $^1$H-$^{13}$C pairs with the experimental $^1$H-$^{13}$C 2D peaks in the HETCOR spectrum, which shows 2D peaks associated with the sidechains at $\delta(^{13}C)$ = 10-40 ppm and $\delta$ ($^1$H) = 1-4 ppm, and the backbone moieties $\delta$ ($^{13}$C) = 110-170 ppm and $\delta$ ($^1$H) = 5-9 ppm (Figure 3d). However, deviations between the DFT-calculated versus experimental chemical shifts are observed for the 2D peaks corresponding to the through-space aromatic-sidechain dipolar interactions as depicted in the blue dashed boxes. Similar trends are observed for Y-series NFAs that showed changes in the local structures with respect to the backbone-sidechain interactions between the crystalline and powder forms.[58] The most important take away from the ssNMR crystallography study is that it allowed us to leverage the Le Bail refinement as a tool to verify the structural compatibility between single crystal and powder in terms of packing motif. This is possible as it allows for shift in the lattice parameters caused by side chains relaxation at elevated temperatures and demonstrated by changes in the backbone-sidechain interactions, while preserving the π-π interactions. By extending the Le Bail analysis to the other NFAs of interest (Figures S5-S8 and Table S3), we obtained useful information about the materials crystallinity:

1. ITIC herringbone polymorph is predominant over the brickwork (Figure S7).
2. m-4TICO presents two brickwork structures but only the un-solvated one is represented in powder (Figure S8).
3. o-IDTBR, IDIC, m-ITIC and 4TICO single crystal packing is preserved in powder (Figure 3a and S6).
4. With the only exception of 4TIC,¶ all the NFAs show several Bragg peaks in the low angle region.
5. All the non-solvated structures undergo a volumetric expansion due to the temperature difference between single crystal and powder experiments. However, we do not exclude that the volumetric reduction observed for the solvated crystals is only apparent, given that a significant volume portion in the single crystal structure is occupied by the solvent, which is not expected to be found in the powder structure.

A summary of the results obtained by Le Bail refinement performed for the different NFAs can be found in Table S3.

**From Powder to NFA Films**

A further intermediate step to approach the NFA packing in the BHJ was to study the molecular organisation in thin films. Here, we expected to see a more compatible unit cell to the one observed in powder rather than in single crystal as we were going towards systems that are presumably composed of many little crystallites with reduced long-range order, cumulative disorder and multiple orientations.[34] Moreover, the temperature held during GIWAXS measurements on film was 300 K as for the powder experiments (XRD and ss-NMR). For convenience, we here report the analysis performed on *o*-IDTBR, while the complete dataset including the other materials can be found in the Supporting Information (pages S10-S19).

GIWAXS data of *o*-IDTBR film with related 1D integration profiles along the in-plane and out-of-plane directions are shown in Figure 4a-b and Table 2. From the q-map two main contributions are visible: a low-angle component, located at $q \approx 3.8$ nm$^{-1}$, which is generally recognised as lamellar peak and is indicative of the separation of the conjugated and aliphatic moieties,[35] and a higher angle feature ($q \approx 18.3$ nm$^{-1}$) which is commonly attributed to π-π stacking.[64] Given the anisotropic nature of these two main diffraction components, we expected a π-π stacking with a preferential face-on crystalline orientation of the *o*-IDTBR domains. To validate our hypothesis, we simulated the GIWAXS pattern of the *o*-IDTBR single crystal structure (Le-Bail refined) oriented along the (4 1 1) direction (Figure 4c), which is nearly parallel to the π-π stacking (4 0 2) and perpendicular to the lamellar (0 1 -1) peak (Figure 4e). The good agreement between the simulated and experimental diffraction data suggests that the *o*-IDTBR packing motif is preserved in film, where the domains adopt a face-on orientation with an in-plane lamellar ordering (0 -1 1) and out-of-plane π-π stacking (4 0 2). The good agreement between our findings with literature[65, 66] clarifies the crystal packing motif and orientation of *o*-IDTBR films.

Some more considerations on the *o*-IDTBR film crystallinity can be done by focussing on the spectral shape of the main diffraction peak (lamellar peak). According to the paracrystalline *g* parameter found for the lamellar peak (Table 2),[34] the film can be classified on the boundary between semi-paracrystalline and amorphous, showing a Crystal Coherence Length (CCL) of ~ 20 nm. For this class of materials, a direct quantification of the crystalline domain size from the CCL is often not possible. According to the nomenclature used in ref.[34], we will refer to the CCL as the spatial extent of the coherently diffracting regions included in the paracrystallites, i.e. column lengths.[34]

To access the NFA domain size (which can be composed of multiple paracrystallites), we investigated the surface morphology by AFM. Figure 4d shows well-defined domain boundaries and a root mean square (RMS) roughness of 6.9 nm. We performed the AFM image segmentation (see pages S20-S25) through watershed algorithm to derive the average domain size and its distribution.[44] From the calculations, we observed an average domain size of 36 nm from the maximum of the peak distribution (inset of Figure 4d and Table 2). This value is higher than the CCL, which confirmed that o-IDTBR domains (visible from the AFM) are composed of multiple paracrystallites, whose column length is determined by XRD (CCL). To get an indication of the domain structural purity, we introduced a parameter (*ϕ*) defined as the ratio (in percent) between the CCL and the domain size obtained by AFM (Table 2).

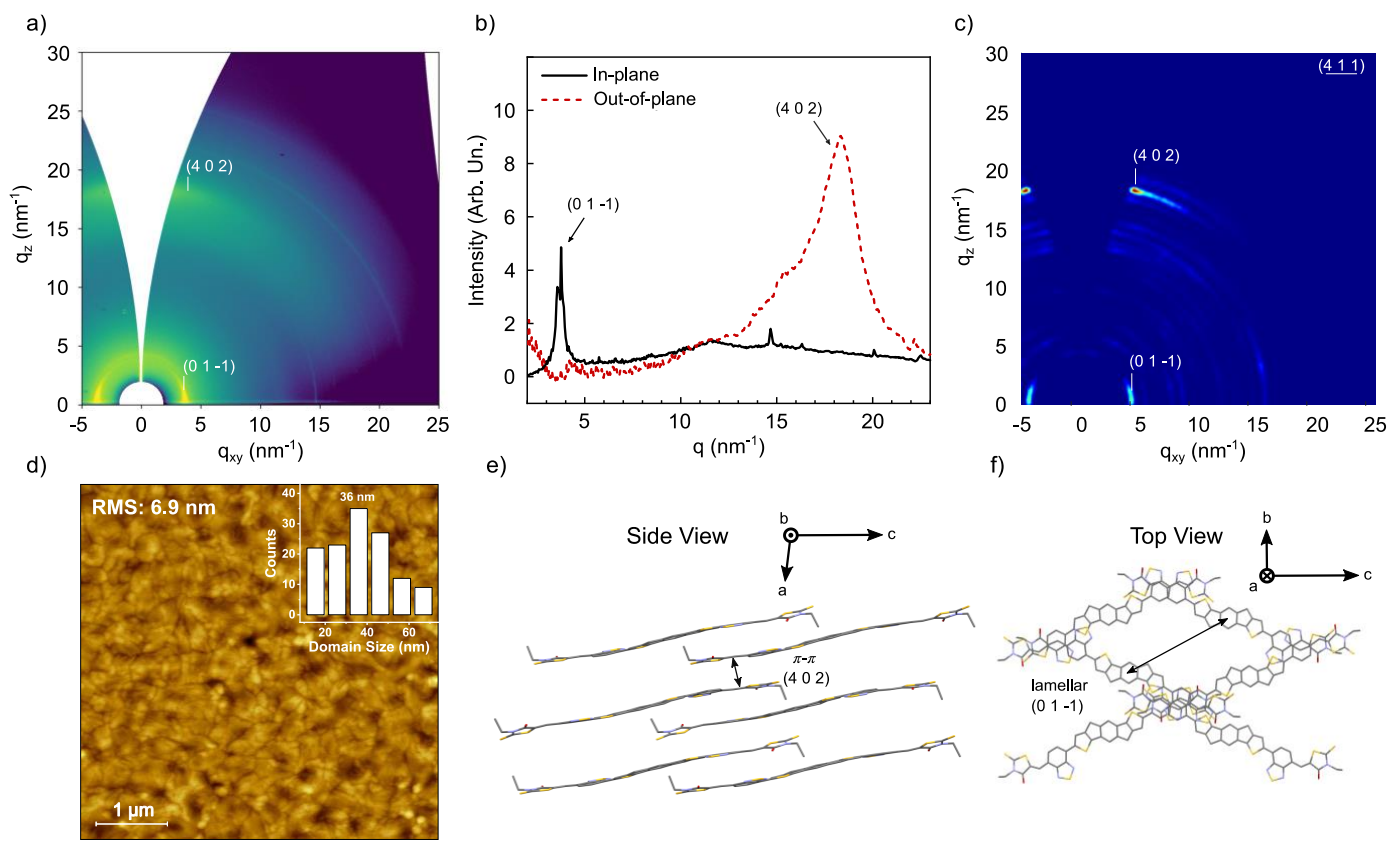

Figure 4. o-IDTBR GIWAXS pattern (a) with in-plane and out-of-plane integration profiles along $q_z \sim 0$ and $q_{xy} \sim 0$, respectively. (b). Simulated GIWAXS pattern of the o-IDTBR unit cell oriented along the (4 1 1) direction. The good agreement with the experimental GIWAXS confirmed a face-on 3D reticular packing motif of *o*-IDTBR in the blend. (c). 5x5 μm AFM image of o-IDTBR film with domain size distribution and average value (inset). Side (e) and top (f) views of the o-IDTBR crystal packing with π-π stacking (4 0 2) and lamellar (0 1-1) peaks.

Analogue characterisation and analysis were performed on the other NFAs of interest for this work (Figure S9-S14 and Tables S4-S9) and some key considerations and results can be summarised as follows:

1. most of the NFAs films showed a well-defined GIWAXS scattering, especially in the lamellar and π-π stacking regions (Figure S9-S12 and Tables S4-S7). These two main features are characterised by a low angular distribution and therefore are indicative of a preferential orientation of the crystalline domains with respect to the substrate.
2. *o*-IDTBR, *m*-4TICO, 4TIC, *m*-ITIC and IDIC crystal lattice simulations yielded a good structural agreement with the powder unit cell obtained by Le Bail refinement (Figures S9-S12a). This proved that a structural continuity in terms of packing motif occurs between powder and films.
3. NFAs with a 3D reticular (4TIC and *o*-IDTBR, see Figures S9b and 5e,f) and 2D brickwork (*m*-ITIC and IDIC, see Figures S11-S12b) crystal packing motifs are involved in a face-on domain orientation, with the *m*-4TICO as only exception ("quasi" edge-on crystal packing, see Figure S10b).
4. ITIC and 4TICO were found to have a 0D-herringbone packing motif in powder. However, due to the lack of multiple Bragg peaks, texturing, and long-range crystallinity (Figures S13, S14 and Tables S8-S9) we could not perform any crystal lattice simulation. However, we do not exclude the presence of small and randomly oriented herringbone column lengths within the domains.
5. The high $g$ parameter found for all the NFAs (> 9.5), prevented us from directly quantifying the crystallite domain size from the lamellar peak shape (FWHM) as the CCL represents the spatial extent of the coherently diffracting regions (Table S4-S9). We therefore estimated the domain size from AFM image segmentation along with an indication for the domain purity ($\phi$).
6. In general, NFAs that form π-π stacking structures (through 2D brickwork or 3D reticular motifs) in single crystals and powder showed the highest crystallinity in films (lowest $g$ parameters and highest CCL, see Tables S4-S9). Conversely, 0D herringbone NFAs (ITIC and 4TICO) provided the highest paracrystalline parameter $g$, highest CCL, and surprisingly among the highest domain purity $\phi$.

**NFAs blended with PBTZT-stat-BDTT-8**

After having characterised the NFA film crystallinity, we extended our investigation on NFA:PBTZT-stat-BDTT-8 blend films, which were used as active layers for the solar cells fabrication (see below). As for the rest of the structural characterisation, we here limit the discussion on PBTZT-stat-BDTT-8:*o*-IDTBR blend while the analysis for the other systems can be found in the Supporting Information.

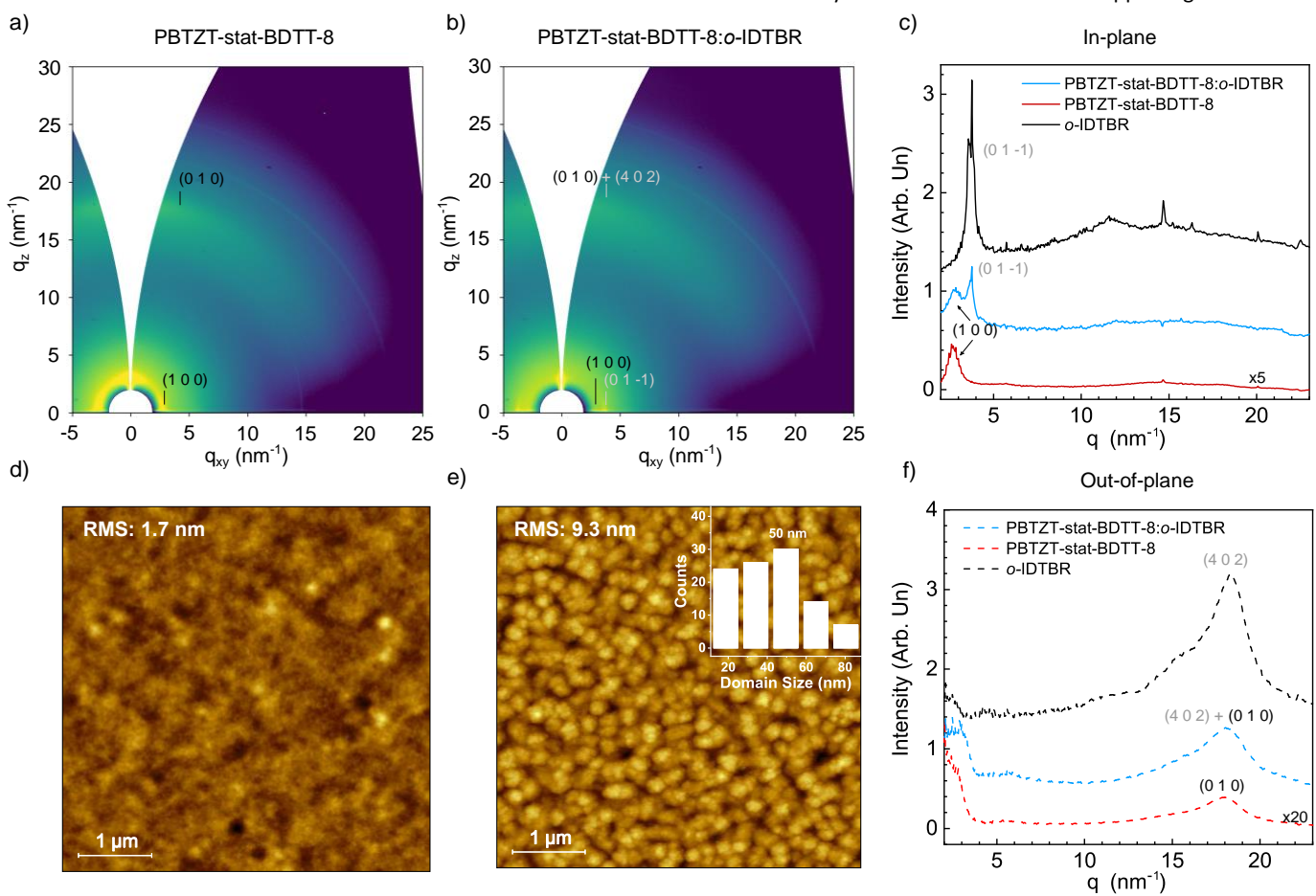

Figure 5. PBTZT-stat-BDTT-8 (a) and PBTZT-stat-BDTT-8:o-IDTBR (b) GIWAXS patterns with in-plane (c) and out-of-plane (f) integration profiles. The NFA features are clearly visible from the blend GIWAXS, which confirmed that *o*-IDTBR maintain the packing motif in blend. 5x5 µm AFM images of PBTZT-stat-BDTT-8 (d) and PBTZT-stat-BDTT-8 :o-IDTBR (e) with domain size distribution and average value (inset).

| Component | Peak | Orientation | q (nm⁻¹) | d (nm) | FWHM (nm⁻¹) | CCL (nm) | g | Domain Size (nm) | φ (%) |
|---|---|---|---|---|---|---|---|---|---|
| | | | *NFA film* | | | | | | |
| o-IDTBR | (0 1 -1) | In-plane | 3.79 | 1.66 | 0.27 | 20.9 | 10.6 | 36 ± 7 | 58.1 |
| o-IDTBR | (4 0 2) | Out-of-plane | 18.35 | 0.34 | 2.22 | 2.5 | 13.9 | - | - |
| | | | *Polymer film* | | | | | | |
| PBTZT-stat-BDTT-8 | (1 0 0) | In-plane | 2.71 | 2.32 | 0.78 | 7.2 | 21.4 | 24 ± 5 | 30.2 |
| PBTZT-stat-BDTT-8 | (0 1 0) | Out-of-plane | 18.05 | 0.35 | 0.75 | 7.5 | 25.7 | - | - |
| | | | *Blend film* | | | | | | |
| o-IDTBR | (0 1 -1) | In-plane | 3.77 | 1.67 | 0.33 | 17.1 | 11.8 | 50 ± 10 | 34.3 |
| PBTZT-stat-BDTT-8 | (1 0 0) | In-plane | 2.80 | 2.24 | 1.58 | 3.6 | 30.0 | - | - |
| PBTZT-stat-BDTT-8, o-IDTBR | (0 1 0) + (4 0 2) | Out-of-plane | 18.14 | 0.35 | 2.80 | 2.0 | 15.7 | - | - |

Table 2 Crystallographic information of the main peaks observed by GIWAXS on o-IDTBR, PBTZT-stat-BDTT-8 and PBTZT-stat-BDTT-8:o-IDTBR films. AFM domain size and purity are also shown. The domain purity parameter (φ) is defined as the ratio (in percent) between the CCL and the domain size obtained by AFM.

The PBTZT-stat-BDTT-8 polymer GIWAXS pattern is shown in Figure 5a, where a (1 0 0) lamellar reflection is located at $q \approx 2.7$ nm⁻¹ and the (0 1 0) π-π stacking feature at $q \approx 18.0$ nm⁻¹. The integration profiles suggest a prevalent in-plane orientation of the (1 0 0) feature and an out-of-plane direction of the (0 1 0) (Figure 5c, f). A slight face-on crystalline orientation of the PBTZT-stat-BDTT-8 was previously reported, along with its smooth surface morphology with low RMS (Figure 5d).[36]

GIWAXS data and 1D profiles of PBTZT-stat-BDTT-8:o-IDTBR blend are shown in Figure 5b. A broad π-π stacking feature is located at $q \approx 18.1$ nm⁻¹ along the out-of-plane (Figure 5d), which can arise from both the NFA and the polymer due to the spectral overlap in the $q$ range. Therefore, we focus on the lamellar features of o-IDTBR and PBTZT-stat-BDTT-8 as indicative of the distinct material ordering in the blend given that they can be distinguished from the in-plane profiles (Figure 5c). The o-IDTBR (0 1 -1) lamellar peak is located here at $q \approx 3.8$ nm⁻¹, meaning that the NFA crystalline ordering in the blend is preserved with a similar lattice spacing and crystal packing.

A remarkable difference is reported for the spectral shape of the NFA lamellar peak as it is characterised by an increased FWHM, indicating a reduced long-range ordering (lower CCL) of the o-IDTBR domains in the blend when compared to pure o-IDTBR film, resulting in higher $g$. In addition to this, the increased domain size calculated from the AFM image (Figure 5e), which implies a lower degree of domain purity (φ) of the NFA in the blend (Table 2).

The analysis for the other NFAs can be found in the Supporting Information (Figures S9-S14 and Tables S4-S9). However, the main conclusions can be outlined as follows:

1. The NFA crystallinity in PBTZT-stat-BDTT-8:NFA blends presented broader and slightly shifted (towards lower $q$) lamellar peaks. As a result, the NFA domains in the blend are characterised by reduced crystallinity (lower CCL, higher $g$) and relaxed lamellar packing with respect to the bare NFA film. Furthermore, the presence of the polymer is also affecting the domain purity (φ), which is reduced for most of the blends with respect to the films made of NFAs.

2. NFAs with 3D-reticular (o-IDTBR and 4TIC) and 2D-brickwork (m-4TICO, IDIC and m-ITIC) arrangements in pure NFA films, preserved their packing motif and texturing in NFA:PBTZT-stat-BDTT-8 blends (Figure 5 and pages S10-S17).

3. NFAs that formed 0D herringbone structures in single crystal and powder phases (ITIC and 4TICO), showed the lowest crystallinity (highest $g$ parameter and lowest CCL calculated on the NFA lamellar peak) and poor texturing among the series of NFA:PBTZT-stat-BDTT-8 blends (pages S18-S19). Our results are in good agreement with a recent report, where the importance of the π-π stacking interaction energy to preserve the NFA packing motif in the blend films is highlighted.[67]

**Solar Cells Characteristics**

To investigate the role of the NFA packing motif and crystallinity on the charge transport properties and performance of OSCs, we fabricated inverted architecture devices (Figure 6a). The NFAs of interest were tested with PBTZT-stat-BDTT-8 active layers and optimised with respect to the choice of the solvent, casting temperature, and post-annealing treatment (Table S10). The energy levels of the different NFAs with respect to the donor polymer are shown in Figure 6b, along with the UV-Vis spectra of each active layer used (figure 6c). The JV curves delivering the highest PCE are shown in Figure 6d, and the JV characteristics are listed in Table 3 (dark JV curves in Figure S24). Surprisingly, 4TICO and ITIC were among the best performing NFAs in terms of maximum and average PCE obtained, despite having the lowest crystallinity as indicated by the highest $g$ parameter and lowest CCL (Table 3). Moreover, the NFA crystal packing motif does not seem to have a direct impact on performance (Table 3): 3D reticular packing NFAs such as o-IDTBR and 4TIC reached a maximum PCE of 6.6 and 5.9 while m-4TICO (2D brickwork packing motif) was the least performing NFA (5.3% PCE). Interestingly, NFAs with a 0D herringbone packing motif in single crystal (4TICO and ITIC) delivered the highest performance (8.1 % and 7.2 %, respectively). Nevertheless, we still expected a strong interplay between the active layer crystallinity, NFA packing, and charge transport properties, which may have an impact on the solar cells performance and in particular the FF.[68, 69] Thus, we performed photo-CELIV experiments to determine the charge mobility and the

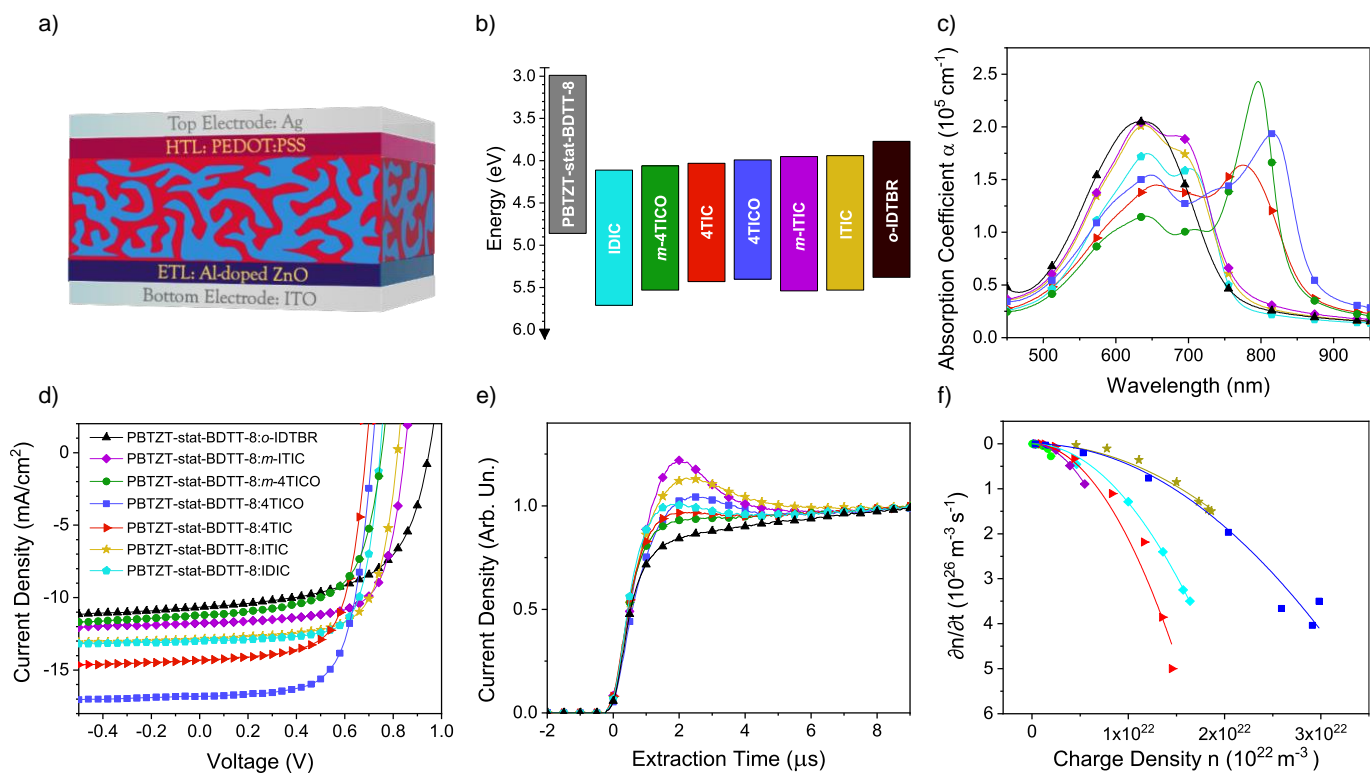

Figure 6. Device architecture with 80 ± 5 nm thick active layers (a) and energy levels determined by Air Photoemission Spectroscopy (see experimental section) (b) of PBTZT-stat-BDTT-8 and the different NFAs used for the solar cells' fabrication. UV-Vis of the different active layers (c). Characterisation of solar cells: J-V characteristics under illumination (dark curves are plotted in Figure S17) (d), photo-CELIV curves (e) and bimolecular recombination coefficients (f). Legends of panels c-f) are shared and shown in d).

bimolecular recombination coefficient for each NFA:PBTZT-stat-BDTT-8 blend (Figure 6e,f and Table 3).¶ ¶.[46, 47, 70-75]

With regards to the mobility, we observed a remarkable correlation between the NFA lamellar CCL in the blend and the charge mobility (Figure 7a). The lowest mobility for ITIC and 4TICO blends is related to their poor crystallinity (low CCL and high $g$ parameter) detected in the pure NFA (Tables S4-S9) and blend films (Table 3). The 0D packing nature might be another disadvantage for efficient long-range transport. The relationship between charge mobility and NFA crystallinity was further investigated by MIS-CELIV measurements, performed on electron-injecting devices. The electron mobility determined by MIS-CELIV matched the values obtained via photo-CELIV (Figure S23). Since the photo-CELIV current is dominated by the species with higher mobility,[76] we could attribute the photo-CELIV mobility to electrons. This lead to the conclusion that the electron mobility is clearly dependent on the NFA crystallinity, as expressed by the CCL and $g$ parameter. However, the mobility determined from both photo-CELIV and MIS-CELIV did not dominate the solar cell performance, as seen from FF and PCE (Table 3). This result encouraged us to investigate the bimolecular recombination and its possible implications in the device performance.

We derived the bimolecular recombination coefficient by photo-CELIV, exploring a purely quadratic dependence between recombination and charge carrier density (Figure 6f).[47] The model provided a good fit with the sweep-out of free charges at different time delays, (Figure 6f) and the bimolecular recombination coefficient ($\beta_{exp}$) was derived according to the equation (2) used in ref.,[47] and is shown in Table 3. Interestingly, we find the lowest recombination coefficients for the solar cells made of ITIC and 4TICO blends, which delivered the highest performance and among the highest FF, highlighting the importance of the bimolecular recombination on the device parameters.[36, 77] Despite their lower crystallinity, active layers made of ITIC and 4TICO resulted in higher domain purity ($\phi$). Conversely, NFAs characterised by higher CCL and lower $g$, such as IDIC and $m$-4TICO, provided among the lowest domain purity and highest bimolecular recombination coefficient. Overall, a correlation is found between the NFA domain purity and the bimolecular recombination coefficient (Figure 7b).

As mentioned above, the $\phi$-parameter compares the spatial extent of the NFA ordered regions in the domain (i.e., column lengths, derived from the CCL), with the domain size obtained from AFM images and can be calculated as follows: $\phi = CCL\ (nm)/Domain\ size\ (nm) \times 100$). A blend film with low domain purity ($\phi$) can be understood as formed by domains with a larger relative fraction of regions with an amorphous or mixed nature that prompt recombination.[78] Assessing domain purity via the $\phi$-parameter is easy-to-access compared to more sophisticated methods such as resonant soft X-ray scattering (RSoXS).[79, 80]

Table 3 Solar cells characteristics for the different NFAs combined with PBTZT-stat-BDTT-8. Results for the best solar cell in terms of PCE (average and standard deviation over a minimum of 10 devices) are shown for each active layer. Device mobility and bimolecular recombination coefficients are also listed along the other parameters related to the film crystallinity ($g$, CCL, packing), morphology (RMS, domain size) and domain purity ($\phi$). The energy bandgap of the blend is also shown ($E_g$).

| Active layer | PCE (%) | FF (%) | $V_{OC}$ (mV) | $J_{SC}$ (mA·cm$^{-2}$) | $\mu$ (cm$^2$ V$^{-1}$ s$^{-1}$) | $\beta_{exp}$ (m$^3$ s$^{-1}$) | $g$ | CCL (nm) | Dom. size (nm) | $\varphi$ (%) | RMS (nm) | Packing | $E_g$ (meV) |
|---|---|---|---|---|---|---|---|---|---|---|---|---|---|
| PBTZT-stat-BDTT-8:4TICO | 8.1 (7.5 ± 0.3) | 70.9 (68.5 ± 1.2) | 725 (717 ± 4) | 16.8 (15.3 ± 0.7) | 5.2 ± 0.5 x10$^{-5}$ | 4.6 ± 2.3 x 10$^{-18}$ | 22.4 | 5.8 | 12.1 ± 2 | 47.7 | 1.4 | 0D | 870 |
| PBTZT-stat-BDTT-8:ITIC | 7.2 (6.8 ± 0.2) | 69.3 (67.5 ± 1.1) | 820 (816 ± 3) | 12.9 (12.4 ± 0.4) | 5.7 ± 0.6 x10$^{-5}$ | 4.2 ± 2.1 x 10$^{-18}$ | 21.0 | 5.7 | 9.6 ± 2 | 59.4 | 1.2 | 0D | 920 |
| PBTZT-stat-BDTT-8:IDIC | 7.0 (6.8 ± 0.2) | 73.0 (71.2 ± 1.7) | 750 (745 ± 2) | 13.4 (12.8 ± 0.4) | 1.4 ± 0.1 x10$^{-4}$ | 1.3 ± 0.6 x 10$^{-17}$ | 15.4 | 10.5 | 41.1 ± 8 | 25.5 | 5.1 | 2D | 750 |
| PBTZT-stat-BDTT-8:m-ITIC | 6.9 (6.6 ± 0.3) | 69.7 (67.7 ± 1.6) | 850 (837 ± 7) | 12.2 (11.7 ± 0.5) | 8.7 ± 0.9 x10$^{-5}$ | 2.6 ± 1.3 x 10$^{-17}$ | 16.3 | 8.4 | 51.5 ± 10 | 16.3 | 6.7 | 2D | 910 |
| PBTZT-stat-BDTT-8:4TIC | 6.6 (6.4 ± 0.2) | 67.5 (65.6 ± 1.1) | 685 (685 ± 5) | 14.3 (14.2 ± 0.3) | 1.1 ± 0.1 x10$^{-4}$ | 2.1 ± 1.0 x 10$^{-17}$ | 14.1 | 9.0 | 29.6 ± 6 | 30.3 | 2.7 | 3D | 830 |
| PBTZT-stat-BDTT-8:o-IDTBR | 5.9 (5.6 ± 0.2) | 58.7 (57.6 ± 0.6) | 950 (944 ± 3) | 10.8 (10.3 ± 0.4) | N/A | N/A | 11.8 | 17.1 | 50.0 ± 10 | 34.3 | 10.1 | 3D | 1090 |
| PBTZT-stat-BDTT-8:m-4TICO | 5.3 (4.6 ± 0.6) | 63.7 (61.5 ± 1.6) | 765 (755 ± 6) | 11.2 (9.9 ± 1.1) | 1.4 ± 0.1 x10$^{-4}$ | 1.2 ± 0.6 x 10$^{-16}$ | 14.9 | 9.6 | 68.5 ± 14 | 14.0 | 11.7 | 2D | 800 |

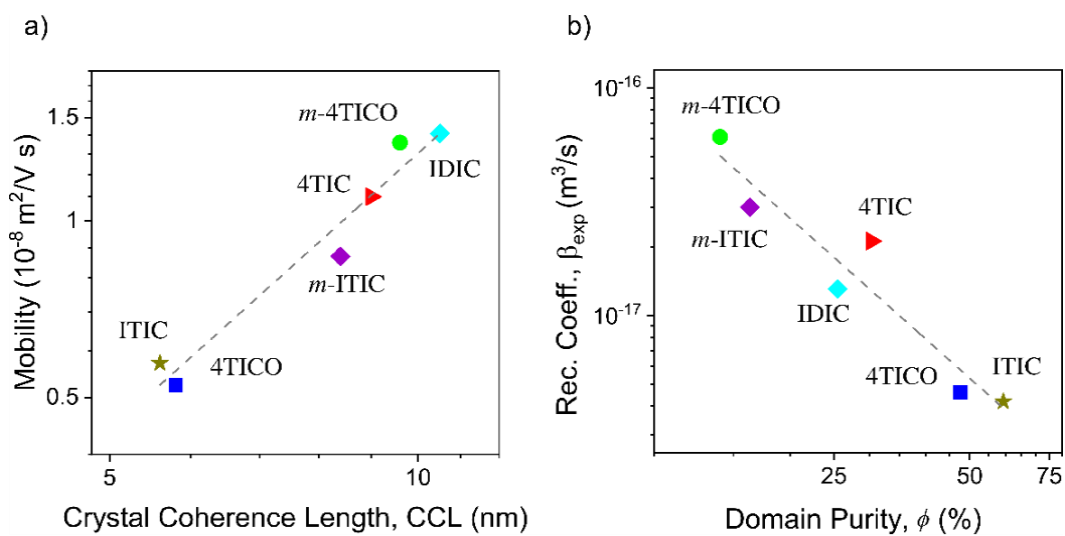

Figure 7 a) Mobility is plotted versus the NFA crystal coherence length in the blend films. A linear fit of the data is represented with a dashed grey line and indicates a correlation between CCL and electron mobility. b) Bimolecular recombination coefficient determined by photo-CELIV in relation to the NFA domain structural coherency in the blend. A linear fit is represented with a dashed grey line and indicates a correlation between the bimolecular recombination coefficient and the domain purity.

To summarise the influence of the NFA crystallinity ($g$, CCL), packing motif and morphology (domain size, RMS and $\phi$) on the solar cell parameters (PCE,[#] FF, $V_{OC}$ and $J_{SC}$) and charge transport properties ($\mu$ and $\beta_{exp}$), we built a multivariable cross-correlation map with the most relevant parameters[##] (Figure 8). While, strictly speaking, the cross-correlation map tests for linear correlation, data that is correlated in a non-linear fashion will still result in high (absolute) values. As such the map may serve as semi-quantitative tool to assess correlations in complex multivariable systems, where not all dependencies are fully understood physically.

With regards to the initial motivation of our study on the role of the NFA crystal packing, we didn't observe a clear correlation with the device performance parameters. However, an enhanced propensity for NFAs with increasing directionality of π-π stacking to form crystalline domains in the blend is evident by the high Pearson correlation coefficient ($r$) between the packing motif with $g$ and CCL, 0.97 and 0.76, respectively. Thus, the directionality of the π-π stacking directly promotes the NFA crystallinity in blends ($g$ and CCL), which is, in turn, favouring the electron mobility ($r$ coefficient of $\mu$ is 0.97 and -0.88 with CCL and $g$. It is worth to mention a recent work, in which a long exciton diffusion lifetime was observed in systems with enhanced π-π stacking systems and crystallinity.[67] The tendence to obtain higher mobilities in organic solar cells by increasing the materials crystallinity is generally acknowledged.[34, 81-83] In addition to this, a lack of a direct correlation between packing motif and performance was also recently reported, although the identification of the NFA packing motif in the blend was based on a visual examination of the GIWAXS pattern and 0D herringbone systems were not included in the study .[67]

Interestingly, we found that the bulk heterojunction morphology is having a bigger impact on performance. In particular, active layers forming big domains at the surface have low domain purity ($r$ = -0.9), which is a good correlator for the bimolecular recombination coefficient (Figure 7b-8 and S25). $\beta_{exp}$ is, in turn, a first-tier correlator to $J_{SC}$ and FF ($r$ is -0.79 and -0.81, respectively) and the best correlating factor for the solar cell performance ($r$ = 0.94). The primary importance of the bimolecular recombination on the device performance and its relation with the domain purity was also observed in literature.[84]

According the cross-correlation analysis, $V_{OC}$ is generally not dependent on the film crystallinity and morphology. The energy bandgap of the blend $E_G$, determined with the difference between HOMO of the polymer and the LUMO of the NFA (Figure 6b), is the only significant correlator to $V_{OC}$. The last result agrees with the general knowledge about the relation between $V_{OC}$ and the energetics of the donor and the acceptor used in the blend.[85-87]

## Conclusions

We have studied how NFAs crystal packing evolves from single crystals to the bulk heterojunction of a solar cell. Given the complexity of unambiguously determine the NFA packing motif in an active layer, arising when moving from ideal systems, i.e., millimetre-size single crystals, to the most complex BHJ morphology, we employed a step-by-step structural analysis. The first step involved ss-NMR crystallography and powder XRD, which helped us to leverage the Le Bail refinement as a quick and effective tool to verify the structural compatibility between single crystal and powder samples. Then, we combined GIWAXS, crystal lattice simulations and AFM to systematically identify the NFA packing motif in the bare NFA films and blends with the PBTZT-stat-BDTT-8 polymer, and to derive key parameters to describe the material crystallinity (CCL and $g$ parameter) and morphology (RMS, domain size and domain purity, $\phi$). Finally, we investigated the influence of those key

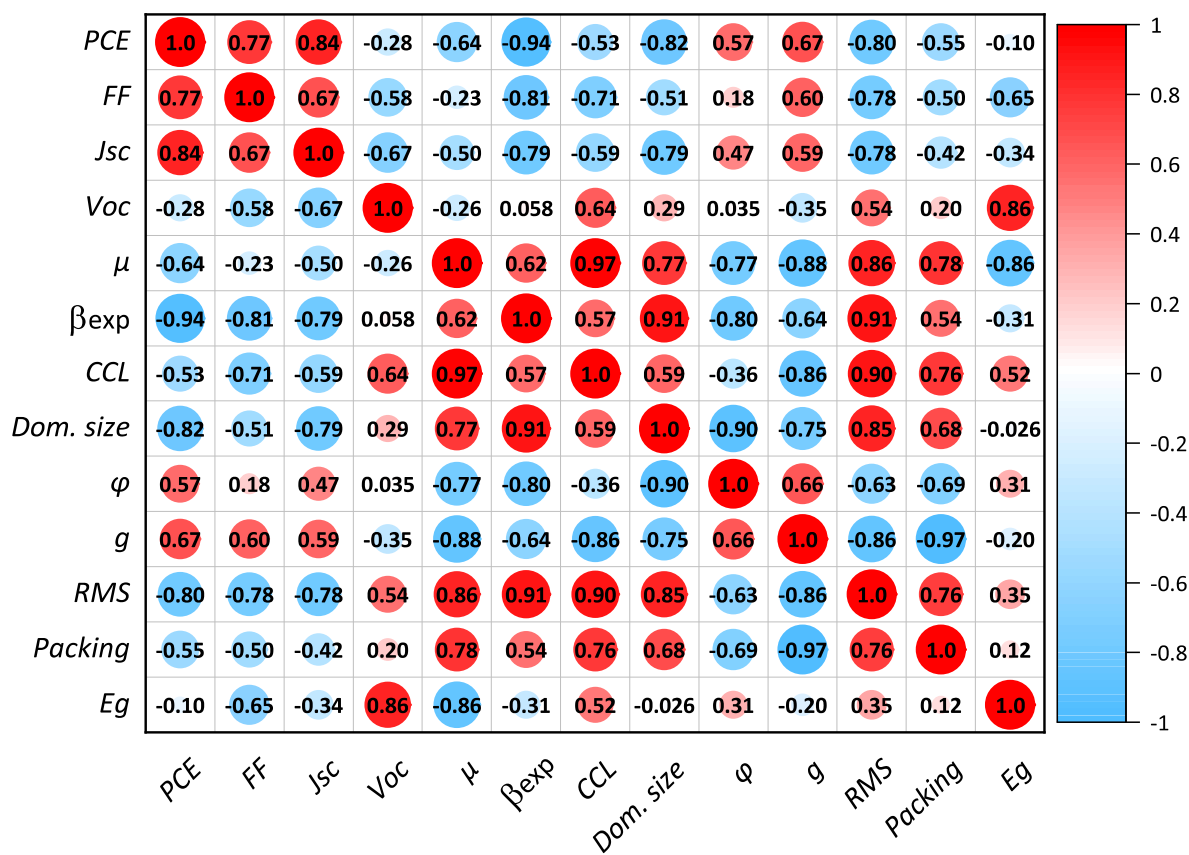

Figure 8. Multivariable cross-correlation map between solar cell characteristics and crystallinity/morphology parameters.

structural parameters on the solar cell performance and charge transport properties.

Our main findings are:

1. NFA packing motifs largely track from single crystals to the thin-film blend.
2. Compounds that crystallise easily as single crystal also show high crystallinity in the blend films. For instance, we found that the poor propensity of ITIC to form single crystals (as indicated by the multiple unsuccessful trials to grow single crystals)[18] also translates into a low blend crystallinity (high $g$ and low CCL).
3. NFAs with higher π-π stacking dimensionality showed an increased propensity to form crystalline films (low $g$ and high CCL) in both NFA films and blends.
4. Despite our initial expectations, the NFA packing motif does not directly correlate with the solar cell performance parameters.
5. NFAs with high film crystallinity (low $g$, high CCL) provided higher electron mobility. However, the mobility is not the dominating factor for device performance. At the same time, we found no correlation between the NFA crystallinity in the blend and the bimolecular recombination coefficient ($\beta_{exp}$).
6. The bimolecular recombination coefficient ($\beta_{exp}$) is found to be the main factor influencing FF and $J_{SC}$. Systems with low $\beta_{exp}$ reported the highest performance. For instance, blend with lower NFA crystallinity (4TICO and ITIC) delivered the highest performance and lowest bimolecular recombination despite the lowest electron mobility.
7. Domain purity stood out as an interesting design target, to limit the bimolecular recombination and obtain high efficiencies in organic solar cells. A better understanding of the influence of molecular properties on domain purity is needed.

A high domain purity could be targeted through chemical design that aims at limiting void space within the unit cell while also managing the solubility/miscibility of the donor-acceptor pairing to control BHJ formation and intermolecular interactions.[36, 88] This could in theory be targeted by the design of space filling yet flexible sidechains to increase the NFA rotational freedom and prevent NFA crystallisation that might induce an excessive phase segregation.[37] Alternatively, thermal annealing can also be used as a handle to tune the morphology of kinetically trapped systems,[89, 90] allowing the formation of a controlled BHJ morphology characterised by domain with high structural purity.[36, 89, 91] This was observed for NFAs that can rearrange their structures undergoing an endothermic transition (glass or liquid crystalline) during post-annealing process (ITIC and 4TICO). Conversely, systems with shorter or sterically locked sidechains may promote the formation of crystalline domains even before any thermal treatment (4TIC, m-4TICO, m-ITIC).[36] These domains tend to excessively phase segregate upon annealing without improving their domain purity.

## Conflicts of interest

Merck KGaA provided the PBTZT-stat-BDTT-8 polymer and the 4TICO NFA.

## Acknowledgements

P. Mondelli and M. Riede acknowledge the European Union's Horizon 2020 research and innovation programme under Marie Skłodowska Curie Grant agreement no. 722651 (SEPOMO) for the support in the realization of this work. P. Kaienburg acknowledges funding from the Global Challenges Research Fund (GCRF) through STFC, START project ST/R002754/1; and from EPSRC for a Postdoctoral Fellowship EP/V035770/1. GIWAXS experiments were performed at NCD-SWEET beamline at ALBA synchrotron with the collaboration of ALBA staff (proposal 2019093873).

## Footnotes

‡Cu-K$\alpha$ X-Ray penetration depth on our films is ~ 11 nm for incident angles $\alpha$ = 0.11° (assuming the same critical angle of $\theta$c = 0.17°).

‡‡The active layer thickness was determined by using a Veeco DEKTAK 150 surface profilometer.

§Single crystal structures are measured under a continuous flow produced by liquid nitrogen at 100 K, while powder diffraction is measured at 300 K.

§§We assume in this first step that the unit cells angles ($\alpha$, $\beta$ and $\gamma$) are not varying from the single crystal unit cell.

¶4TIC powder XRD data quality didn't allow to perform Le Bail analysis due to poor scattering (Figure S6).

¶¶We did not detect a meaningful signal from o-IDTBR:PBTZT-stat-BDTT-8 to extract the mobility and recombination coefficient. We believe this is due to a combination of factors, among which the lowest shunt resistance leading to the lowest FF observed among all the different devices.

#Although the PCE is a linear combination of FF, $V_{OC}$ and $J_{SC}$, we included it in the cross-correlation analysis for convenience.

##We assigned a constant value for each motif which is representative of the dimensionality of the π-π stacking: "0" for 0D herringbone, "2" for 2D brickwork and "3" for 3D reticular.

# Supporting Information:

# Understanding the Role of Non-Fullerene Acceptors Crystallinity on the Charge Transport Properties and Performance of Organic Solar Cells

Pierluigi Mondelli,*[a,d] Pascal Kaienburg,[a] Francesco Silvestri,[b] Rebecca Scatena,[a] Claire Welton,[c] Martine Grandjean,[d] Vincent Lemaur,[e] Eduardo Solano,[f] Mathias Nyman,[g] Peter N. Horton,[h] Simon J. Coles,[h] Esther Barrena,[b] Moritz Riede,[a] Paolo Radaelli,[a] David Beljonne,[e] G. N. Manjunatha Reddy[c] and Graham Morse[d] (Authors are listed in an arbitrary order and list is temporary)

[a] Clarendon Laboratory, University of Oxford, Parks Road, Oxford, OX1 3PU, United Kingdom. E-mail: pierluigi.mondelli@gmail.com

[b] Institut de Ciència de Materials de Barcelona, ICMAB-CSIC, Campus UAB, 08193 Bellaterra, Spain

[c] University of Lille, CNRS, Centrale Lille, Univ. Artois, UMR 8181- UCCS - Unité de Catalyse et Chimie du Solide, F-59000 Lille, France

[d] Merck Chemicals Ltd, Chilworth Technical Centre, University Parkway, Southampton, SO16 7QD, United Kingdom

[e] Laboratory for Chemistry of Novel Materials, University of Mons, Place du Parc, 20, 7000 Mons (Belgium)

[f] ALBA Synchrotron Light Source, NCD-SWEET beamline, Cerdanyola del Vallès, 08290 Spain

[g] Physics, Faculty of Science and Engineering, Åbo Akademi University, 20500 Turku, Finland

[h] EPSRC Crystallographic Service, Department of Chemistry, University of Southampton, Highfield, SO17 1BJ, UK

## Table of Contents

## 1. Solid-state NMR analysis of o-IDTBR powder

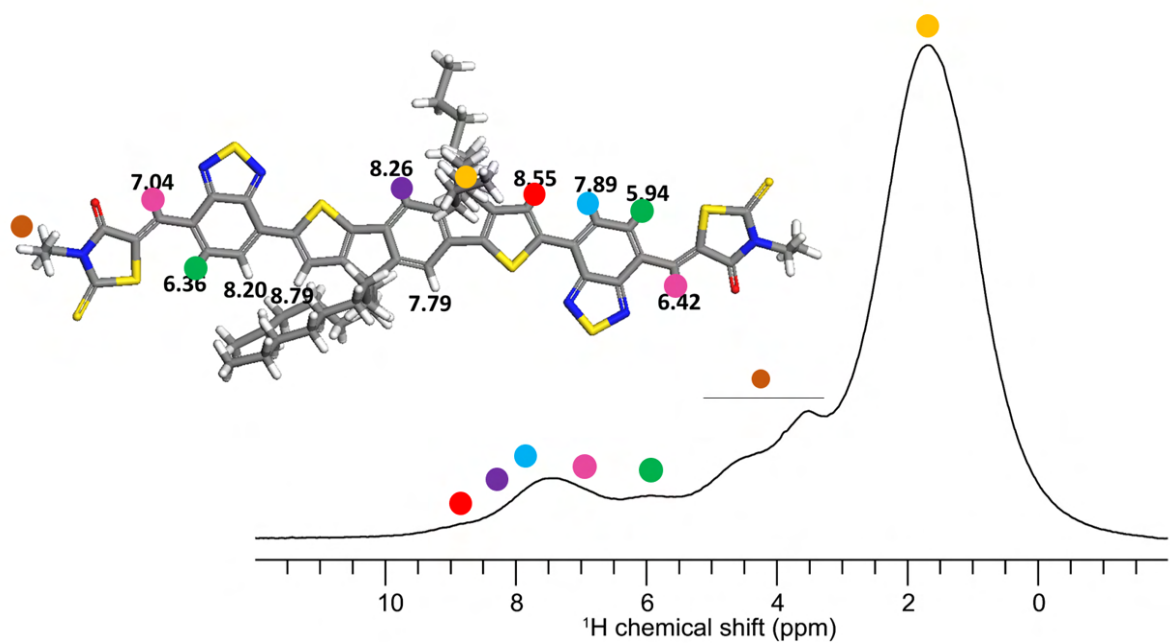

**Figure S1.** 1D solid-state ¹H NMR spectrum of o-IDTBR powder acquired at 18.8 T (¹H = 800 MHz) with 50 kHz MAS. Peak assignments are colour coded as depicted in the crystal structure shown in the inset.

**Figure S1** presents ¹H MAS NMR spectra of O-IDTBR powder, whereby the ¹H peaks are color coded as depicted in the schematic structure figure of O-IDTBR shown on top and the isotropic chemical shifts calculated by GIPAW-DFT approach are overlaid. The broad ¹H peak centered at ~1.7 ppm is due to the branched alkyls sidechains attached to the aromatic core. The different distributions of ¹H peaks in the 3-5 ppm range are due to the -NCH$_2$CH$_3$ moieties attached to the terminal 3-ethyl-2-thioxothiazolidin-4-one groups. In the aromatic region, the peak at ~6 ppm is due to the protons in the benzothiadiazole (BDT) groups (green dots), and the low ppm chemical shift value of these BDT protons can be ascribed to the ring current effects caused by the partial overlap of the aromatic rings. The peak at ~7 ppm can be attributed to the -CH moieties bridging the BDT and 3-ethyl-2-thioxothiazolidin-4-one groups as depicted in the magenta dot. The overlapped peaks in the range between 7.5 to 8.5 ppm are due to the protons in the BDT groups facing towards thiophene (T) groups (blue dot) and the protons in the central benzene ring (purple dots). The weak intensity peak at ~8.8 ppm is due to thiophene protons (red dot).

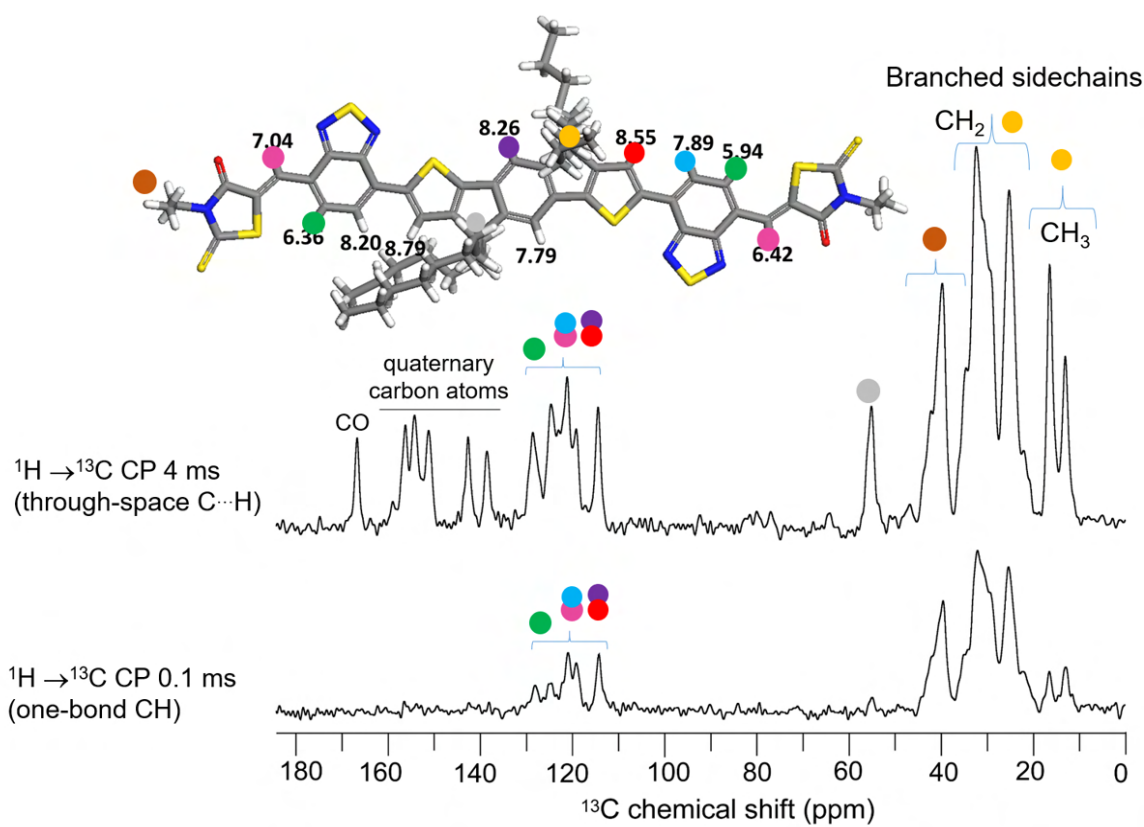

**Figure S2.** 1D solid-state $^{13}C^1$ CP-MAS NMR spectra of o-IDTBR powder acquired with 0.1 (bottom) and 3 ms (top) CP contact time. All spectra were acquired at 18.8 T and 50 kHz MAS. Peak assignments are colour coded as depicted in the crystal structure shown in the inset.

For O-IDTBR powder, the $^1H→^{13}C$ cross-polarization (CP)-MAS NMR spectra of o-IDTBR were acquired with 0.1 ms and 3 ms of CP contact time and compared. CP-MAS based experiments exploit the enhancement of the $^{13}C$ signal intensities that are increased by transfer of $^1H$ spin-polarization by the adjacent protons, according to the strengths of the $^1H$-$^{13}C$ dipole-dipole couplings. The isotropic $^{13}C$ chemical shifts are color coded as presented in the spectrum, the CH$_3$ peaks are well resolved in 10-20 ppm range and the CH$_2$ moieties produce peaks in 20-37 ppm range. The peaks around 40 ppm are due to the -NCH$_2$CH$_3$ groups of 3-ethyl-2-thioxothiazolidin-4-one, and the peak centered at ~55 ppm can be assigned to the quaternary carbon atom bearing the branched alkyl sidechains. In the aromatic region, the one-bond CH moieties produce peaks in the 110-130 ppm; the broad peak in the 110-115 range originates from the protonated carbon atoms in the benzene ring that produces a peak at ~112 ppm (purple dot) and the protonated thiophene carbon atoms that give rise to a peak at ~ 115 ppm (red dot). Partially resolved peaks in the 118-122 ppm range are due to the protonated carbon atom (facing towards the T group as depicted by the blue dot) and the -CH moieties (purple dots) bridging the BDT and 3-ethyl-2-thioxothiazolidin-4-one groups. By comparison, a CP-MAS NMR spectrum of o-IDTBR was acquired with 3 ms CP contact time which displayed additional peaks corresponding to the quaternary carbon atoms (top spectrum), signals of which are enhanced by the CP transfer via through-space $^{13}C$-$^1H$ dipolar interactions. Specifically, the $^{13}C$ peak of the quaternary carbon atom bearing the sidechains (gray dot) is enhanced as shown by the peak at ~55 ppm. In addition, the $^{13}C$ signals associated with the quaternary aromatic carbon atoms at 135-160 ppm are enhanced by the CP transfer from the adjacent protons via $^{13}C$-$^1H$ dipolar couplings. The $^{13}C$ peak at ~168 is due to the carbonyl group of the O-IDTBR molecules.

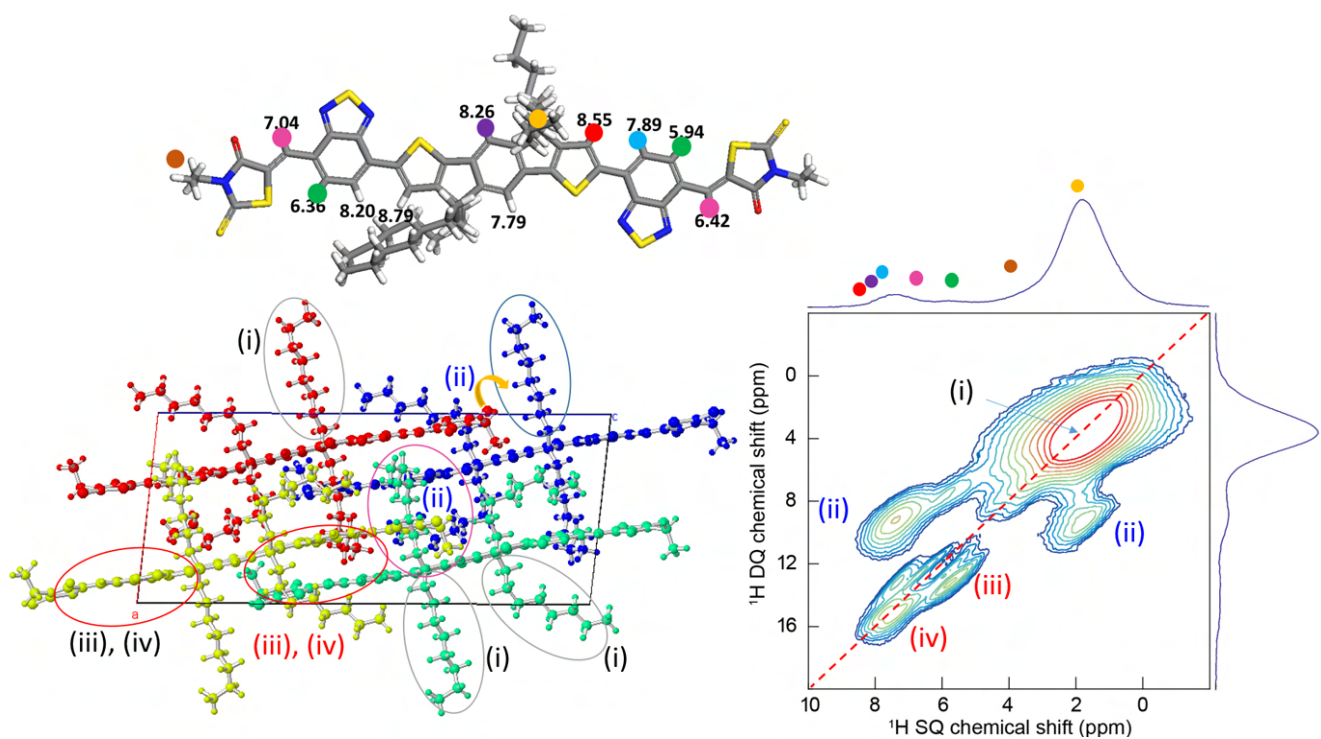

**Figure S3.** 2D solid-state $^1$H-$^1$H NMR spectrum of o-IDTBR powder acquired at 18.8 T ($^1$H = 800 MHz) with 50 kHz MAS. Peak assignments are colour coded as depicted in the crystal structure shown in the inset. $^1$H chemical shift values associated with the aromatic protons are depicted in the GIPAW-DFT geometry optimized structure as shown by the coloured dots.

2D $^1$H-$^1$H double-quantum single-quantum (DQ-SQ) correlation spectrum presented in **Figure S3** displays the DQ peaks in the vertical axis that correspond to $^1$H-$^1$H proximities in less than 5 Å distance. The $^1$H SQ signals are color coded as depicted in the schematic structure figure of O-IDTBR shown on top. In particular, the different aromatic chemical shifts of thiophene (T) and benzothiadiazole (BDT) are due to the different aromatic ring current effects. The inter- and intramolecular $^1$H-$^1$H proximities in the alkyl and aromatic regions (i, ii, iii, and iv) that contribute to the $^1$H DQ peaks are marked by ovals in the crystal structure in the left. The broad DQ peak at ~3.6 ppm (i) is due to the inter-and intramolecular $^1$H-$^1$H proximities in branched sidechains. The DQ peaks at ~7.7 and ~9.4 ppm (ii) are due to the dipolar coupled $^1$H-$^1$H pairs between branched sidechains and aromatic groups (T and BDT). The DQ peaks in the aromatic region (iii and iv) are due to the intramolecular $^1$H-$^1$H proximities between the aromatic moieties, whereby the DQ peak 12.1 ppm is attributable to intramolecular $^1$H-$^1$H dipolar interactions between the BDT proton and C-H protons in the bridged position (green and magenta dots).The DQ peak at 13.1 ppm is due to the intramolecular $^1$H-$^1$H dipolar interactions between the BDT protons (blue and green dots), and the peak at 15.2 ppm is ascribed to the intramolecular $^1$H-$^1$H dipolar interactions between the BDT and T protons (blue and red dots) of O-IDTBR molecules. The different packing interactions that contribute to the DQ peaks are marked by ovals in the crystal structure depicted in the left.

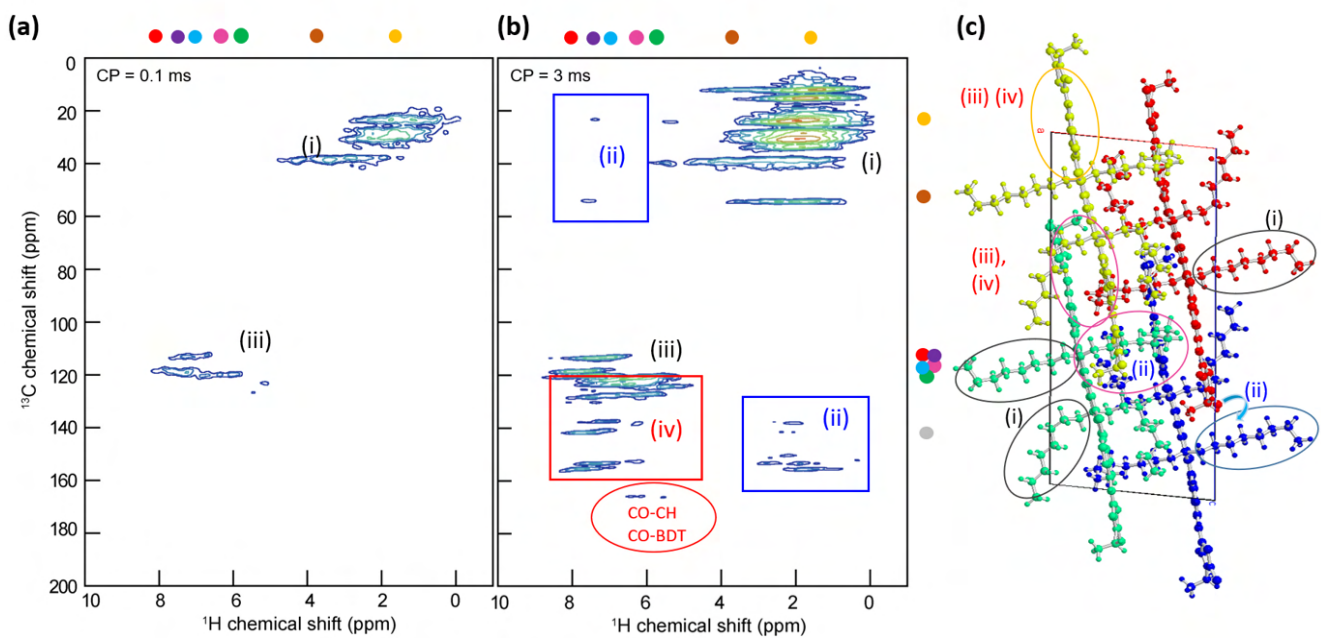

**Figure S4.** 2D solid-state $^1$H-$^{13}$C heteronuclear correlation (HETCOR) spectrum of o-IDTBR powder acquired with (a) 0.1 ms and (b) 3 ms CP contact time. Peak assignments are color coded as depicted in the crystal structure shown in (c). All spectra were acquired at 18.8 T ($^1$H = 800 MHz) with 50 kHz MAS.

**Figure S4** presents 2D HETCOR spectra acquired with 0.1 ms and 3 ms CP contact times in order to detect 2D peaks that originate from directly bonded C-H as well as through-space dipolar coupled C⋯H moieties in o-IDTBR molecules. Inter- and intramolecular C-H proximities corresponding to the 2D peaks are shown in the crystal structure (c). In the spectrum acquired with 0.1 ms CP contact time, the isotropic $^{13}$C chemical shifts at $\delta(^{13}C)$ = 10-45 ppm are due to the directly bonded C-H moieties in alkyl groups (yellow dots), and the isotropic $^{13}$C chemical shifts at $\delta(^{13}C)$ = 110-122 ppm are due to the directly bonded C-H moieties in BDT (blue and green dots), T (red dot), CH moiety at the bridged position (magenta) and benzene ring (purple dots) at the aromatic core. In addition to these, the 2D HETCOR spectrum acquired with 3 ms CP contact time exhibits the 2D peaks corresponds to the through-space inter and intramolecular $^1$H-$^{13}$C dipolar interactions between aliphatic and aromatic groups, as depicted in the boxes and ovals. Specifically, the 2D $^{13}$C-$^1$H correlation peaks associated with the quaternary carbon atoms that are in close proximities to aromatic protons in the core are emerged as depicted in the red color box. The 2D peaks shown in the blue boxes are due to the inter- and intramolecular $^{13}$C-$^1$H dipolar interactions between the alkyl chains and the aromatic moieties. The carbonyl carbon atom is in close proximity to the CH protons (magenta) and intermolecularly with the BDT protons (green dots) that lead to the 2D peaks as depicted in the red oval.

## 2. Powder and Single Crystal X-Ray Diffraction

**Manual calculation of the lattice parameters**

The *o*-IDTBR lattice parameters (*a*, *b* and *c*) for powder sample were initially obtained by solving the reciprocal-space metric tensor for monoclinic structures:

$$\frac{1}{d^2} = \frac{1}{\sin^2\beta}\left(\frac{h^2}{a^2} + \frac{k^2\sin^2\beta}{b^2} + \frac{l^2}{c^2} - \frac{2hl\cos\beta}{ac}\right)$$

where the positions of the (0 1 1), (0 1 2) and (110) were derived from the experimental pattern. The unit cell angles ($\alpha$, $\beta$ and $\gamma$) were assumed to be the same as in single crystal. The results are shown in (Table S1) and compared to the single crystal unit cell and the ones obtained by Le Bail refinement of powder data.

**Table S1.** Evolution of the o-IDTBR unit cell parameters from the known single crystal data to the manual and Le Bail refinements.

| Unit Cell | a (Å) | b (Å) | c (Å) | $\alpha$ (°) | $\beta$ (°) | $\gamma$ (°) | Volume (Å³) |
|---|---|---|---|---|---|---|---|
| Single crystal | 13.7663(2) | 15.81032(17) | 32.7146(3) | 90 | 96.2928(12) | 90 | 7077.43(15) |
| Manual | 14.0646 | 16.0432 | 32.7057 | 90 | 96.293 | 90 | 7335.288 |
| Le Bail | 14.03086 | 16.06125 | 32.68444 | 90 | 96.716 | 90 | 7314.994 |

**Temperature effect on the single crystal unit cell**

The effect of the temperature on the NFA unit cell parameters was explored by performing temperature-dependent single-crystal XRD on IDIC[‡] (Figure S5, Table S2). The experiments have shown how temperature can influence both lattice parameters and angles of IDIC.

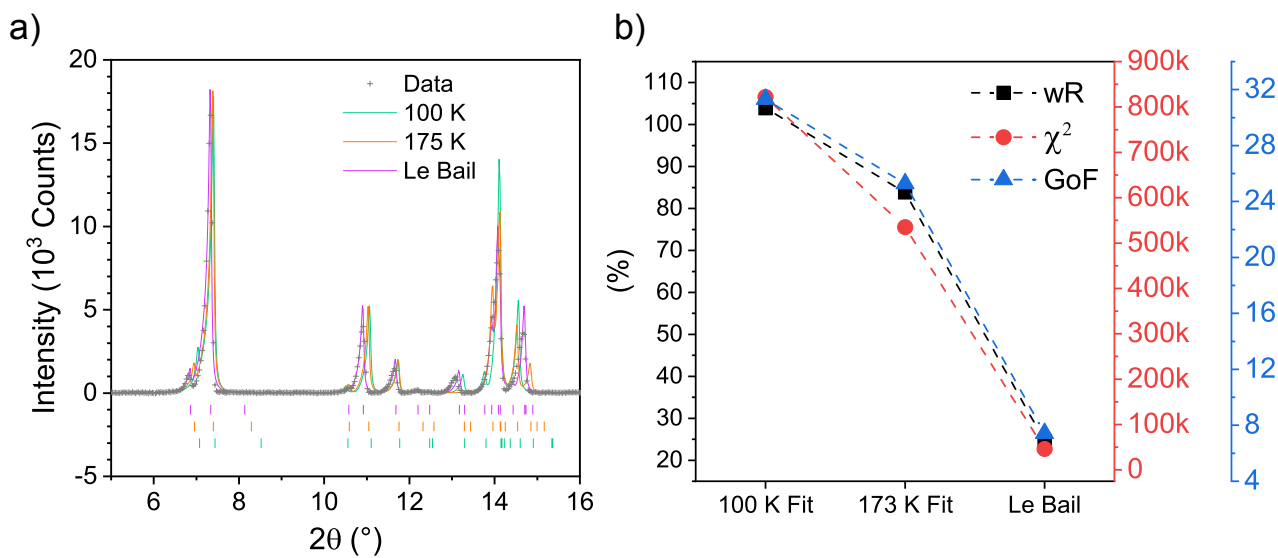

**Figure S5** a) IDIC powder XRD datapoints with Le Bail refinement compared to the simulated powder pattern from the single crystals measured at 100 and 173 K. b) Fitting parameters evolution towards lower values

---

[‡] The IDIC single crystal structures are accessible from the CCDC Database: Deposition Number 2226731 for the measurement taken at 173 K and Deposition Number 2040188 for the dataset taken at 100 K.

**Table S2** IDIC unit cell parameters evolution with temperature

| Structure | Crystal System | Space Group | a (Å) | b (Å) | c (Å) | α (°) | β (°) | γ (°) | Volume (Å³) |
|---|---|---|---|---|---|---|---|---|---|
| 100 K | triclinic | P$\bar{1}$ | 8.6679(4) | 12.5073(7) | 13.5784(6) | 72.096(4) | 75.545(4) | 88.839(4) | 1353.88(12) |
| 173 K | triclinic | P$\bar{1}$ | 8.6379(4) | 12.6391(6) | 13.9041(5) | 70.650(4) | 75.044(4) | 88.148(4) | 1381.35(11) |
| Le Bail | triclinic | P$\bar{1}$ | 8.63994 | 12.80154 | 14.39796 | 69.913 | 74.405 | 88.479 | 1436.737 |

**Le Bail analysis**

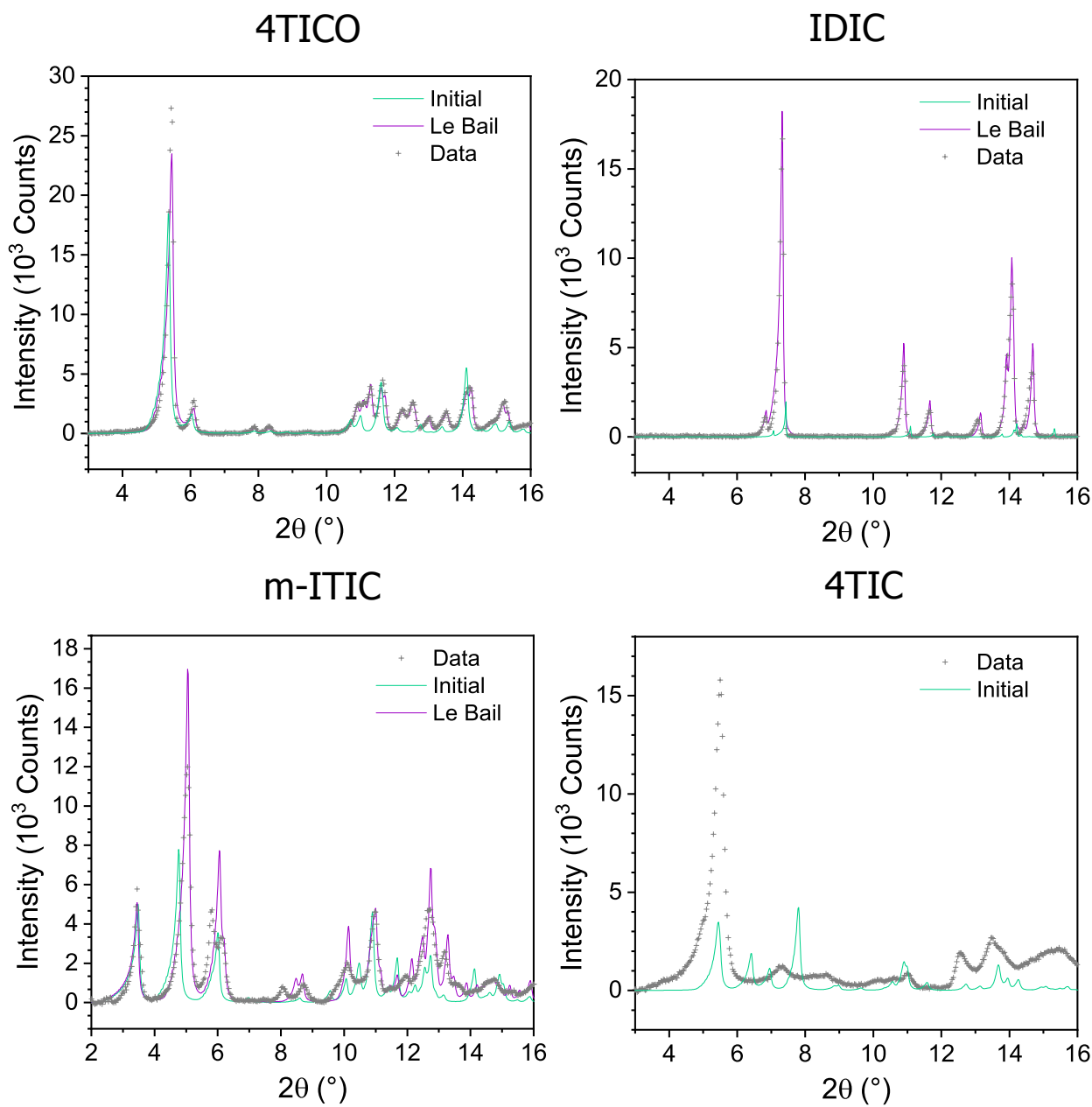

**Figure S6.** Powder XRD datapoints with Le Bail refinement compared to the simulated powder pattern from the single crystal.

# ITIC

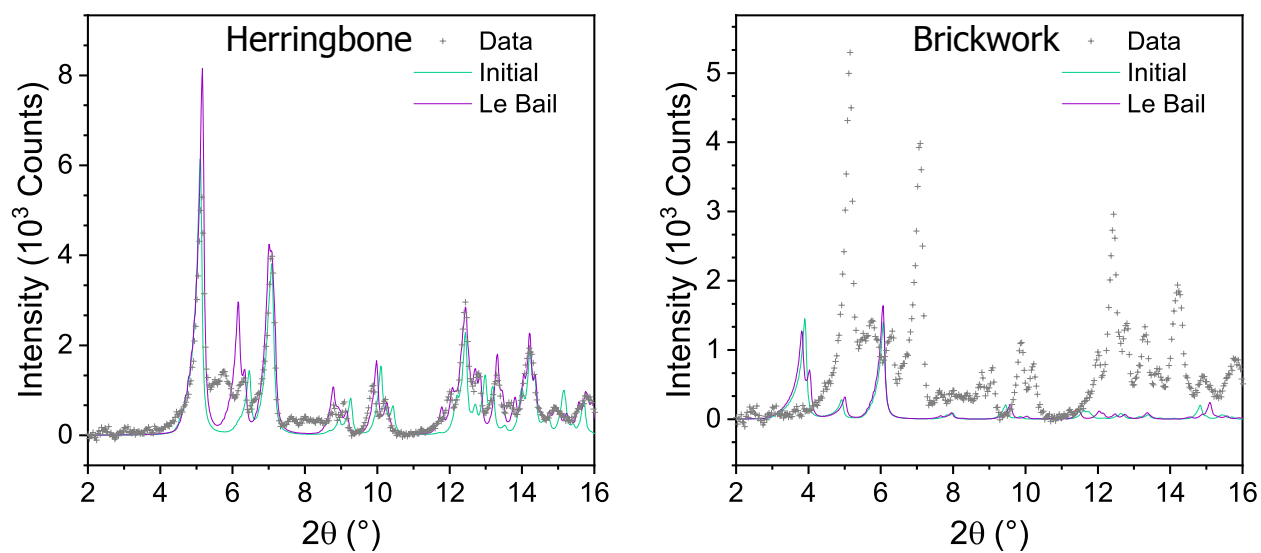

**Figure S7.** Powder XRD datapoints with Le Bail refinement compared to the simulated powder pattern from the single crystal.

# m-4TICO

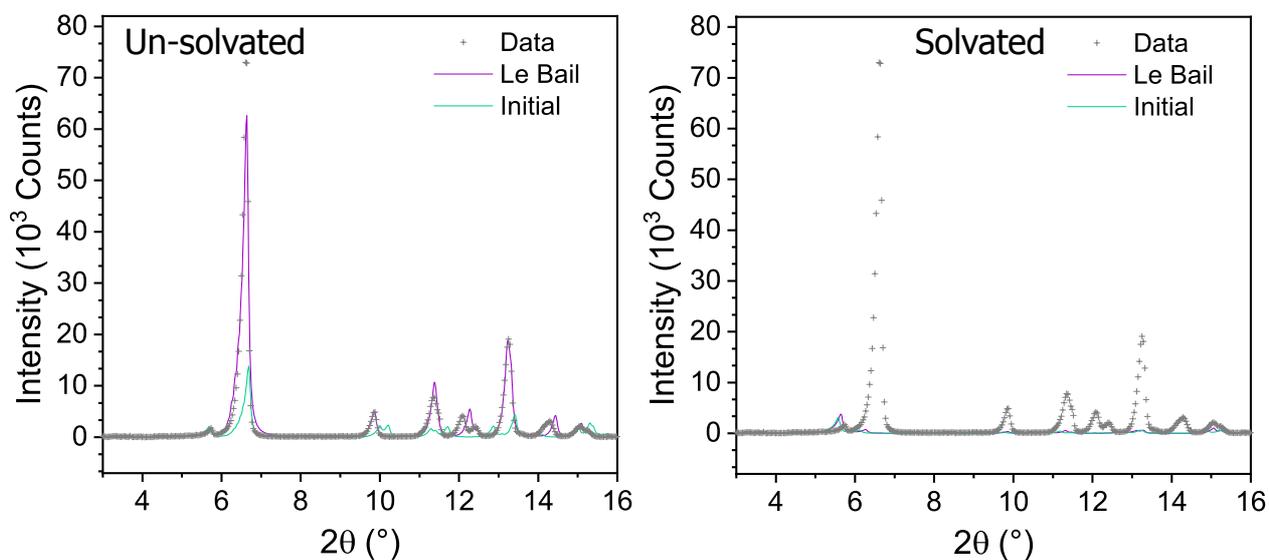

**Figure S8.** Powder XRD datapoints with Le Bail refinement compared to the simulated powder pattern from the single crystal.

**Table S3** NFA unit cell parameters for the known crystal structures in comparison with the parameters obtained from Le Bail refinement.

| CCDC | Molecule | Solvates | a (Å) | b (Å) | c (Å) | α (°) | β (°) | γ (°) | Volume (Å³) | GOF | Chi² | wR |
|---|---|---|---|---|---|---|---|---|---|---|---|---|
| | | | | | *Before Le Bail Refinement* | | | | | | | |
| HEHQUJ01 | ITIC | none | 14.9009(7) | 15.5043(4) | 18.1199(5) | 99.309(2) | 101.541(3) | 108.366(3) | 3777.2(2) | 10.43 | 91145.1 | 24.046 |
| KIZSUK | ITIC | CH₂Br₂ | 8.420(6) | 23.019(17) | 23.126(17) | 101.780(10) | 95.319(10) | 91.105(14) | 4366(5) | 18.24 | 278780.8 | 42.055 |
| VUBJOU | m-ITIC | CHCl₃ | 8.7454(13) | 18.872(2) | 25.2647(18) | 87.770(8) | 88.724(9) | 78.001(12) | 4075.1(9) | 20.51 | 352372 | 38.001 |
| VUBJEK | 4TICO | C₃H₆O | 15.2836(2) | 20.01101(5) | 29.3242(6) | 90 | 90 | 29.3242(6) | 8968.1(3) | 18.95 | 300794.4 | 37.892 |
| FOSPOV | o-IDTBR | none | 13.7663(2) | 15.81032(17) | 32.7146(3) | 90 | 96.2928(12) | 90 | 7077.43(15) | 9.10 | 63293.5 | 31.079 |
| VUBKAH | IDIC | none | 8.6679(4) | 12.5073(7) | 13.5784(6) | 72.096(4) | 75.545(4) | 88.839(4) | 1353.88(12) | 20.93 | 367143.8 | 69.427 |
| N/A | m-4TICO | none | 8.7845(7) | 15.3726(13) | 16.7896(13) | 67.136(7) | 85.678(7) | 79.6630(7) | 2054.98 | 35.82 | 1075472.2 | 62.264 |
| VUBJIO | m-4TICO | CHCl₃ | 8.6526(3) | 16.4878(8) | 18.0435(8) | 114.697(5) | 103.822(4) | 90.890(4) | 2251.45(19) | 40.43 | 1369716.7 | 70.267 |
| YEBKEY | 4TIC | C₇H₈, CH₃OH | 13.969(7) | 17.144(9) | 17.970(10) | 104.668(16) | 109.998(17) | 96.169(14) | 3822.08 | 24.62 | 507853 | 44.652 |
| | | | | | *After Le Bail Refinement* | | | | | | | |
| HEHQUJ01 | ITIC | none | 15.06724 | 15.46733 | 17.89986 | 100.067 | 100.804 | 107.462 | 3788.083 | 7.87 | 51968.4 | 18.157 |
| KIZSUK | ITIC | CH₂Br₂ | 8.57001 | 23.49976 | 22.22479 | 100.617 | 92.928 | 93.867 | 4380.067 | 18.16 | 274453.9 | 41.727 |
| VUBJOU | m-ITIC | CHCl₃ | 8.74588 | 17.77874 | 25.48191 | 88.391 | 89.072 | 78.262 | 3877.599 | 10.96 | 1000667.7 | 20.331 |
| VUBJEK | 4TICO | C₃H₆O | 15.66764 | 19.60675 | 29.04299 | 90 | 90.799 | 90 | 8920.893 | 5.65 | 26766 | 11.303 |
| FOSPOV | o-IDTBR | none | 14.03086 | 16.06125 | 32.78444 | 90 | 96.716 | 90 | 7337.375 | 5.02 | 19267.4 | 17.148 |
| VUBKAH | IDIC | none | 8.67925 | 12.79671 | 14.4997 | 70.019 | 74.42 | 88.239 | 1454.553 | 7.43 | 46096.4 | 24.6 |
| N/A | m-4TICO | none | 9.0868 | 15.7656 | 16.68189 | 67.74 | 85.699 | 79.53 | 2167.014 | 16.70 | 233612.4 | 29.019 |
| VUBJIO | m-4TICO | CHCl₃ | 8.74656 | 16.15453 | 17.81585 | 114.163 | 104.908 | 89.942 | 2203.878 | 40.27 | 1348979.8 | 69.733 |

## 3. Film Morphology by 2D-GIWAXS and AFM

**4TIC**

4TIC 2D-GIWAXS pattern (Figure S6c) is in good agreement with the one obtained in our previous report.[2] We have simulated the 2D-GIWAXS map by using the single crystal unit cell available from literature[3] (Figure S9a) and oriented along the (1 0 0) direction. Such direction is parallel to the π-π stacking (4 -1 -1) and the lamellar (1 0 0) peak, which is representative of a face-on orientation of 4TIC domains with respect to the substrate (Figure S9b). Although the (1 0 0) lamellar contribution is not visible from the simulated data (all crystallites are assumed to be perfectly oriented along the (1 0 0) and therefore are not accessible[4]), experimental and simulated GIWAXS are in good agreement confirming a face-on 3D reticular packing with respect to the substrate. As the 4TIC lamellar feature is still evident from the PBTZT-stat-BDTT-8:4TIC GIWAXS (Figure S9d) and out-of-plane integration profile (Figure S6h), we assume that the NFA is maintaining its crystalline order in blend with a slightly relaxed (1 0 0) periodicity (Table S5) compared to the pure NFA film.

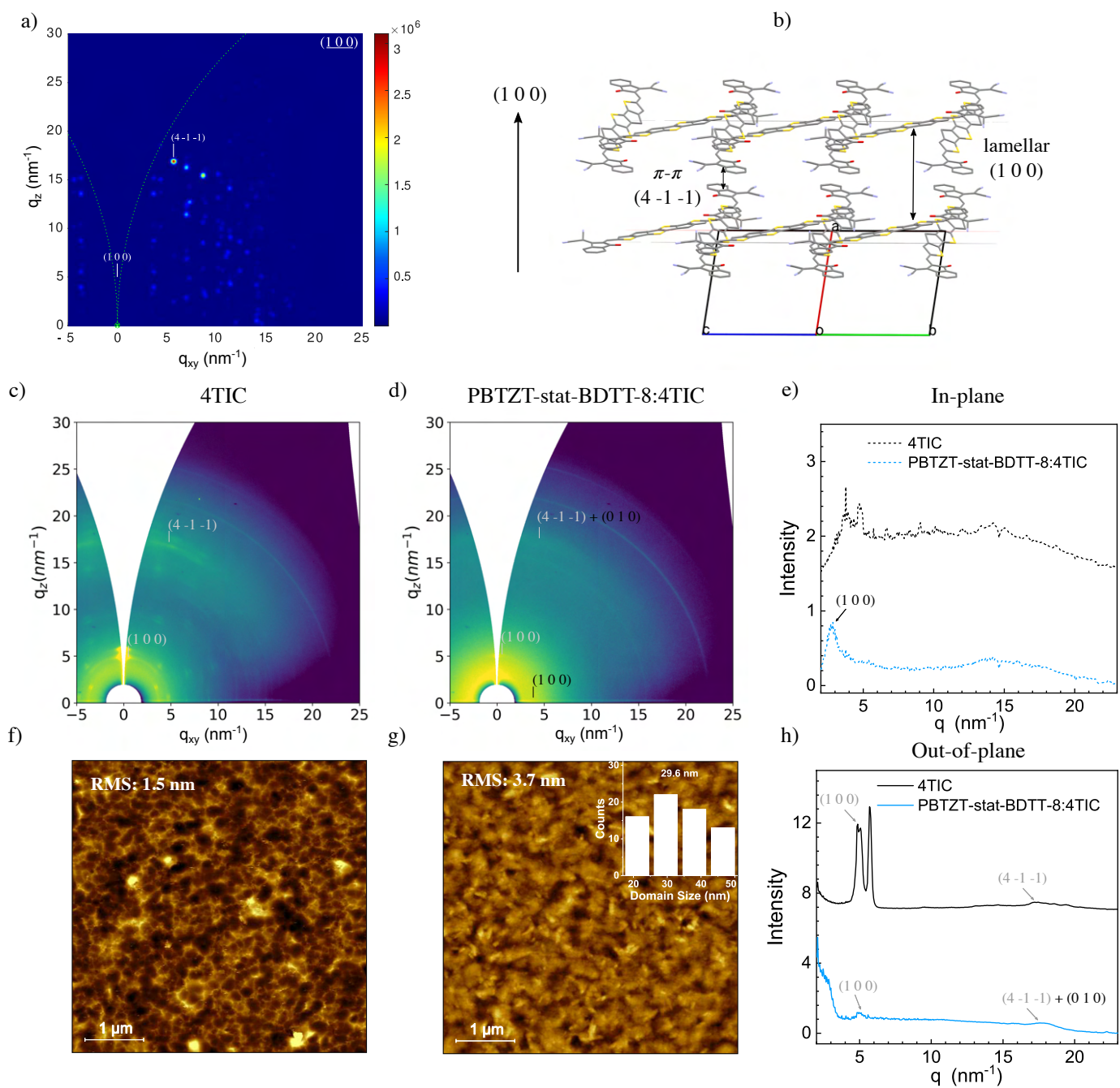

**Figure S9.** Simulated 2D-GIWAXS pattern of the 4TIC unit cell oriented along the (1 0 0) direction (a). Graphical representation of the π-π stacking (4 -1 -1) and lamellar (1 0 0) peaks orientation with respect to the (1 0 0) vector (b). 4TIC (c) and PBTZT-stat-BDTT-8:4TIC (d) GIWAXS pattern with in-plane (e) and out-of-plane (g) integration profiles. 5x5 μm AFM images of 4TIC (f) and PBTZT-stat-BDTT-8:4TIC (g) films with domain size distribution and average value (inset).

**Table S4.** Crystallographic information of the main peaks observed by 2D-GIWAXS on 4TIC and PBTZT-stat-BDTT-8:4TIC films. AFM domain size and domain purity are also shown.

| Component | Peak | Orientation | q (nm$^{-1}$) | d (nm) | FWHM (nm$^{-1}$) | CCL (nm) | g | Domain Size (nm) | φ (%) |
|---|---|---|---|---|---|---|---|---|---|
| *NFA film* | | | | | | | | | |
| 4TIC | (1 0 0) | Out-of-plane | 5.09 | 1.23 | 0.29 | 19.5 | 9.5 | N/A | N/A |
| 4TIC | (4 -1 -1) | Out-of-plane | 17.34 | 0.36 | 1.16 | 4.9 | 10.3 | - | - |
| *Blend film* | | | | | | | | | |
| 4TIC | (1 0 0) | Out-of-Plane | 5.02 | 1.25 | 0.58 | 9.0 | 14.1 | 29.6 | 30.3 |
| PBTZT-stat-BDTT-8 | (1 0 0) | In-plane | 2.75 | 2.28 | 0.87 | 6.5 | 22.4 | - | - |
| PBTZT-stat-BDTT-8, 4TIC | (0 1 0) + (4 -1 -1) | Out-of-plane | 17.81 | 0.35 | 3.00 | 1.9 | 16.4 | - | - |

## m-4TICO

m-4TICO 2D-GIWAXS pattern (Figure S10c) shows a strong Bragg peak along $q_z$ and other less intense features in the q map. To verify whether the NFA arrangement in film is compatible to the phase existing in powder and single crystal, we have simulated the 2D-GIWAXS of the single crystal unit cell (Le Bail refined) and oriented along the (0 1 1) direction (Figure S10a). Such direction is parallel to the lamellar (1 0 0) peak and diagonal with respect to π-π stacking (2 3 2). This is representative of a *quasi*-edge-on orientation of the m-4TICO domains with respect to the substrate (Figure S10b). The good agreement between experimental and simulated GIWAXS confirms a *quasi*-edge-on 2D brickwork packing with respect to the substrate. As the m-4TICO lamellar feature is still evident from the PBTZT-stat-BDTT-8:m-4TICO GIWAXS (Figure S10d) and out-of-plane integration profile (Figure S10h), we assume that the NFA is maintaining its crystalline order in blend with a slightly relaxed (0 1 1) periodicity (Table S5) compared to the pure NFA film.

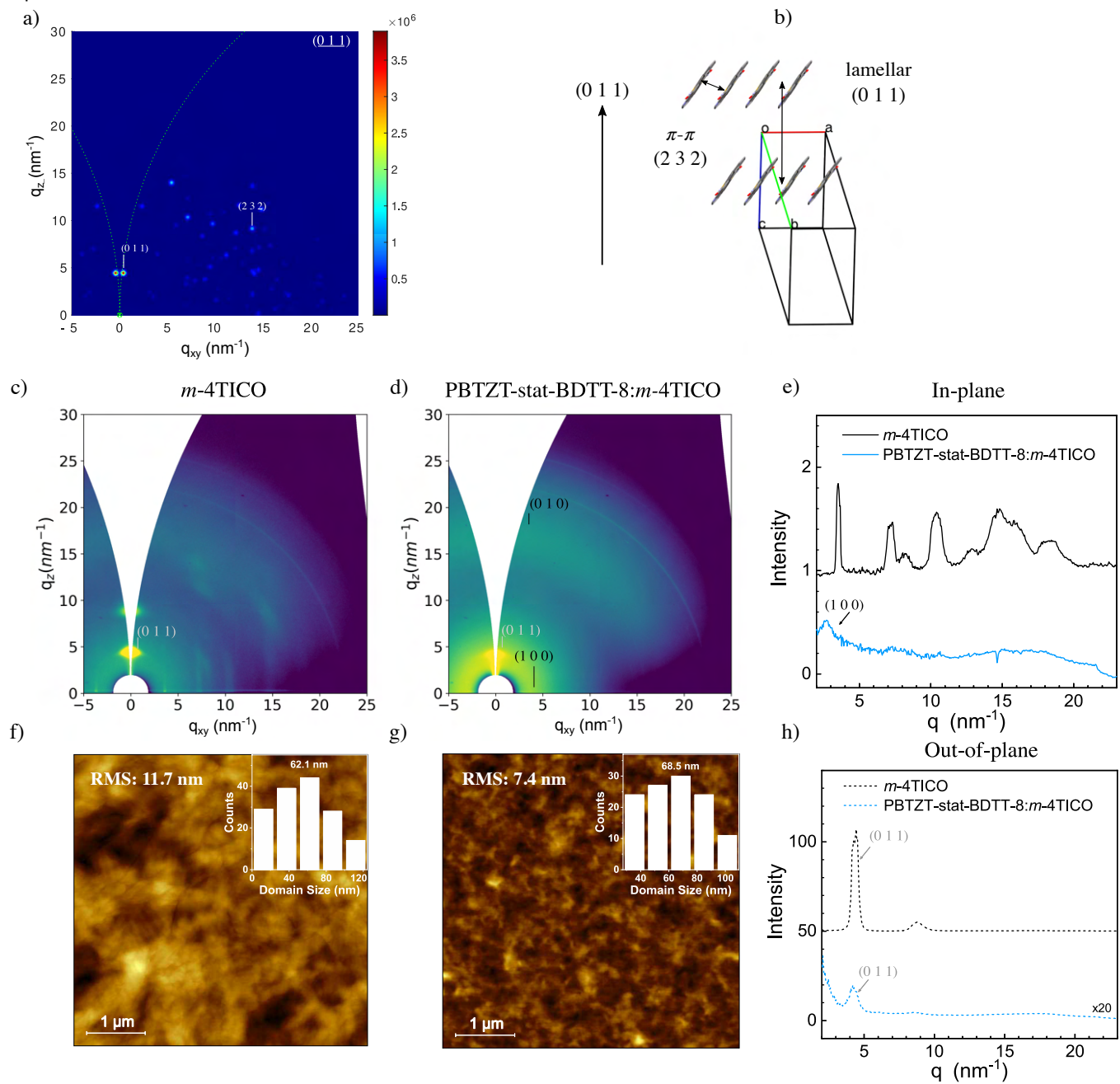

**Figure S10.** Simulated 2D-GIWAXS pattern of the m-4TICO unit cell oriented along the (0 1 1) direction (a). Graphical representation of the π-π stacking (2 3 2) and lamellar (0 1 1) peaks orientation with respect to the (0 1 1) vector (b). m-4TICO (c) and PBTZT-stat-BDTT-8:m-4TICO (d) GIWAXS pattern with in-plane (e) and out-of-plane (f) integration profiles. 5x5 µm AFM images of m-4TICO (f) and PBTZT-stat-BDTT-8:m-4TICO (g) films with domain size distribution and average value (inset).

**Table S5.** Crystallographic information of the main peaks observed by 2D-GIWAXS on m-4TICO and PBTZT-stat-BDTT-8:m-4TICO films. AFM domain size and domain purity are also shown.

| Component | Peak | Orientation | q (nm$^{-1}$) | d (nm) | FWHM (nm$^{-1}$) | CCL (nm) | g | Domain Size (nm) | φ (%) |
|---|---|---|---|---|---|---|---|---|---|
| | | | | *NFA film* | | | | | |
| m-4TICO | (0 1 1) | Out-of-plane | 4.52 | 1.39 | 0.33 | 17.1 | 10.8 | 62.1 | 27.6 |
| | | | | *Blend film* | | | | | |
| m-4TICO | (0 1 1) | Out-of-Plane | 4.25 | 1.48 | 0.59 | 9.6 | 14.9 | 68.5 | 14.0 |
| PBTZT-stat-BDTT-8 | (1 0 0) | In-plane | 2.60 | 2.42 | 1.42 | 4.0 | 29.5 | - | - |
| PBTZT-stat-BDTT-8 | (0 1 0) | Out-of-plane | 17.85 | 0.35 | 5.00 | 1.1 | 21.1 | - | - |

**m-ITIC**

m-ITIC 2D-GIWAXS pattern (Figure S11c) shows a couple of features in the π-π stacking region along $q_z$ and other less intense peaks in the low angle region, both in-plane and out-of-plane. To verify whether the NFA arrangement in film is compatible to the phase existing in powder and single crystal, we have simulated the 2D-GIWAXS of the single crystal unit cell[5] (Le Bail refined) and oriented along the (1 1 1) direction (Figure S11a). Such direction is perpendicular to the lamellar (0 -1 1) peak and parallel with respect to π-π stacking features, (2 3 3) and (2 2 5). This is representative of a face-on orientation of the m-ITIC domains with respect to the substrate (Figure S11b). The good agreement between experimental and simulated GIWAXS confirms a face-on 2D brickwork packing with respect to the substrate. The feature located $q_z \sim 5$ nm$^{-1}$ represents the (1 1 1) reflections but is not evident from the simulation because of its purely perpendicular orientation with respect to the substrate (region not accessible). As the m-ITIC lamellar (0 1 -1) feature is still evident from the PBTZT-stat-BDTT-8:m-ITIC GIWAXS (Figure S11d) and in-plane integration profile (Figure S11h), we assume that the NFA is maintaining its crystalline order in blend with a slightly relaxed lamellar (0 1 -1) and (2 2 5) π-π stacking periodicity (Table S6) compared to the pure NFA film.

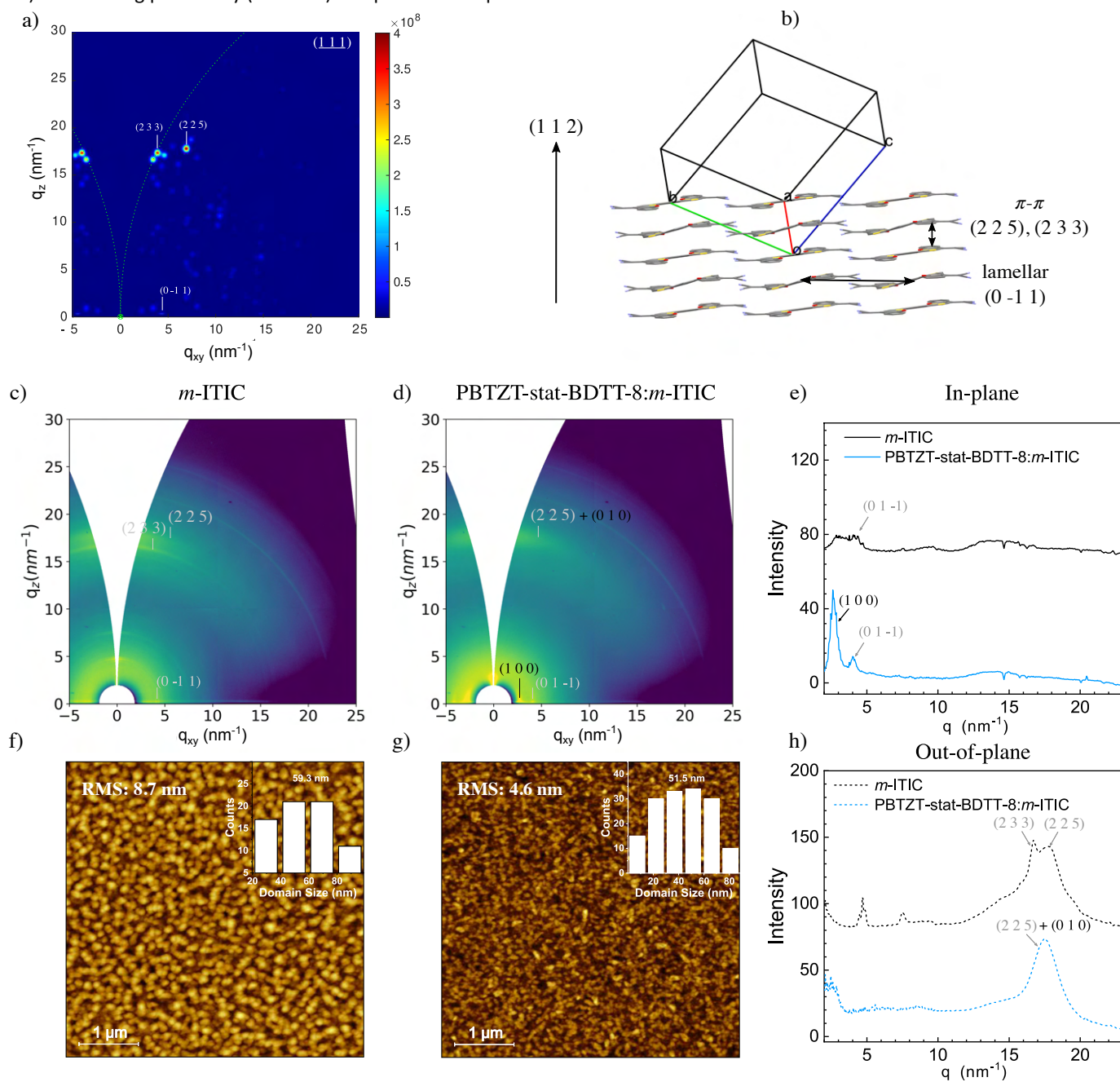

**Figure S11.** Simulated 2D-GIWAXS pattern of the m-ITIC unit cell oriented along the (1 1 1) direction (a). Graphical representation of the π-π stacking, (2 3 3) and (2 2 5), and lamellar (0 1 1) peaks orientation with respect to the (1 1 1) vector (b). m-ITIC (c) and PBTZT-stat-BDTT-8:m-ITIC (d) GIWAXS pattern with in-plane (e) and out-of-plane (h) integration profiles. 5x5 µm AFM images of m-ITIC (f) and PBTZT-stat-BDTT-8:m-ITIC (g) films with domain size distribution and average value (inset).

**Table S6.** Crystallographic information of the main peaks observed by 2D-GIWAXS on m-ITIC and PBTZT-stat-BDTT-8:m-ITIC films. AFM domain size and domain purity are also shown.

| Component | Peak | Orientation | q (nm$^{-1}$) | d (nm) | FWHM (nm$^{-1}$) | CCL (nm) | g | Domain Size (nm) | φ (%) |
|---|---|---|---|---|---|---|---|---|---|
| *NFA film* | | | | | | | | | |
| m-ITIC | (0 1 -1) | In-Plane | 4.09 | 1.54 | 0.43 | 13.1 | 8.6 | 59.3 | 22.2 |
| m-ITIC | (2 2 5) | Out-of-plane | 17.75 | 0.35 | 2.41 | 2.3 | 14.7 | - | - |
| *Blend film* | | | | | | | | | |
| m-ITIC | (0 1 -1) | In-Plane | 4.03 | 1.56 | 0.67 | 8.4 | 16.3 | 51.5 | 16.3 |
| PBTZT-stat-BDTT-8 | (1 0 0) | In-plane | 2.67 | 2.35 | 0.55 | 10.2 | 18.1 | - | - |
| PBTZT-stat-BDTT-8:m-ITIC | (0 1 0) + (2 2 5) | Out-of-plane | 17.54 | 0.36 | 2.30 | 2.5 | 14.4 | - | - |

**IDIC**

IDIC 2D-GIWAXS pattern (Figure S12c) shows a couple of features in the π-π stacking region along $q_z$ and multiple peaks in the low angle region among multiple directions. To verify whether the NFA arrangement in film is compatible to the phase existing in powder and single crystal, we have simulated the 2D-GIWAXS of the single crystal unit cell[5] (Le Bail refined) and oriented along the (2 2 3) direction (Figure S12a). Such direction is nearly perpendicular to the lamellar (0 1 0) peak and parallel with respect to the (2 1 3) π-π stacking feature, which is representative of a face-on orientation of the m-ITIC domains with respect to the substrate (Figure S12b). The partial agreement between experimental and simulated GIWAXS suggests a possible competition between two different polymorphs. The first one represented by the (0 1 0) and (2 1 3) features with a face-on 2D brickwork packing with respect to the substrate, while a second one identified by the (0 1 0)' and (2 1 3)' peaks. We hypothesise for the second polymorph to be still characterised by a face-on packing motif (2D or 3D) because of the presence of an intense out-of-plane (2 1 3)' π-π stacking feature and a (0 1 0)' lamellar peak developing along the in-plane direction. This second polymorph is becoming predominant in PBTZT-stat-BDTT-8:IDIC blend as visible from the GIWAXS data (Figure S12d), integration profiles (Figure S12e,h) and Table S7. As the IDIC lamellar (0 1 0)' feature is still evident and distinguishable from the polymer, we assume that the NFA is still crystalline in the blend with a face-on 2D/3D crystal packing motif.

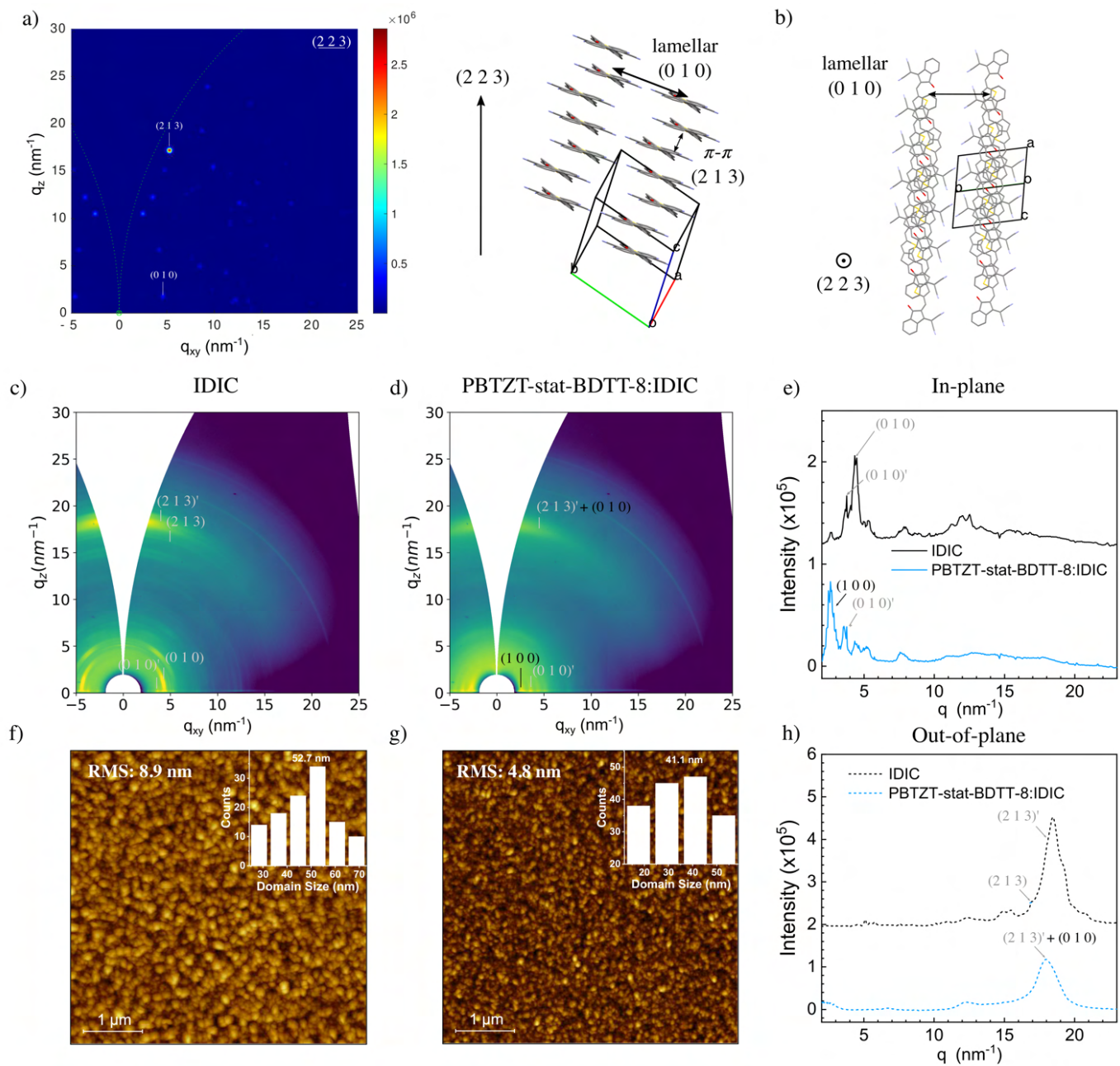

**Figure S12.** Simulated 2D-GIWAXS pattern of the IDIC unit cell (Le Bail refined) oriented along the (2 2 3) direction (a). Graphical representation of the (2 1 3) π-π stacking and (0 1 0) lamellar peaks orientation with respect to the (2 2 3) vector (b). IDIC (c) and PBTZT-stat-BDTT-8:IDIC (d) GIWAXS pattern with in-plane (e) and out-of-plane (h) integration profiles. 5x5 µm AFM images of IDIC (f) and PBTZT-stat-BDTT-8:IDIC (g) films with domain size distribution and average value (inset).

**Table S7.** Crystallographic information of the main peaks observed by 2D-GIWAXS on IDIC and PBTZT-stat-BDTT-8:IDIC films. AFM domain size and domain purity are also shown.

| Component | Peak | Orientation | q (nm$^{-1}$) | d (nm) | FWHM (nm$^{-1}$) | CCL (nm) | g | Domain Size (nm) | φ (%) |
|---|---|---|---|---|---|---|---|---|---|
| *NFA film* | | | | | | | | | |
| IDIC | (0 1 0) | In-Plane | 4.32 | 1.45 | 0.26 | 21.7 | 9.8 | 52.7 | 41.2 |
| IDIC | (2 1 3) | Out-of-plane | 18.46 | 0.34 | 1.44 | 3.9 | 11.1 | - | - |
| *Blend film* | | | | | | | | | |
| IDIC | (0 1 0)' | In-Plane | 3.60 | 1.74 | 0.54 | 10.5 | 15.4 | 41.1 | 25.5 |
| PBTZT-stat-BDTT-8 | (1 0 0) | In-plane | 2.62 | 2.40 | 0.59 | 9.6 | 18.9 | - | - |
| PBTZT-stat-BDTT-8:m-ITIC | (0 1 0) + (2 1 3)' | Out-of-plane | 18.05 | 0.35 | 1.79 | 3.1 | 12.6 | - | - |

## 4TICO

4TICO 2D-GIWAXS pattern (Figure S13a) is characterised by a weak crystallinity as it only presents two main features, both featuring a lack of domain ordering along any specific direction. For this reason, it is not convenient to perform a side-by-side comparison with the single crystal structure.[5] However, a low angle (1 0 0) and a high angle (0 1 0) diffraction rings can still be distinguished (Figure S13a,b) and used for the evaluation of the NFA crystallinity in films (Table S8). As regards the PBTZT-stat-BDTT-8:4TICO blend, we could differentiate between the 4TICO and PBTZT-stat-BDTT-8 lamellar reflections since the latter is typically characterised by in-plane scattering in the low angle region (Figure 5). We therefore conclude that the NFA is weakly crystalline in the blend given the higher FWHM of the observed reflections with lack of texturing.

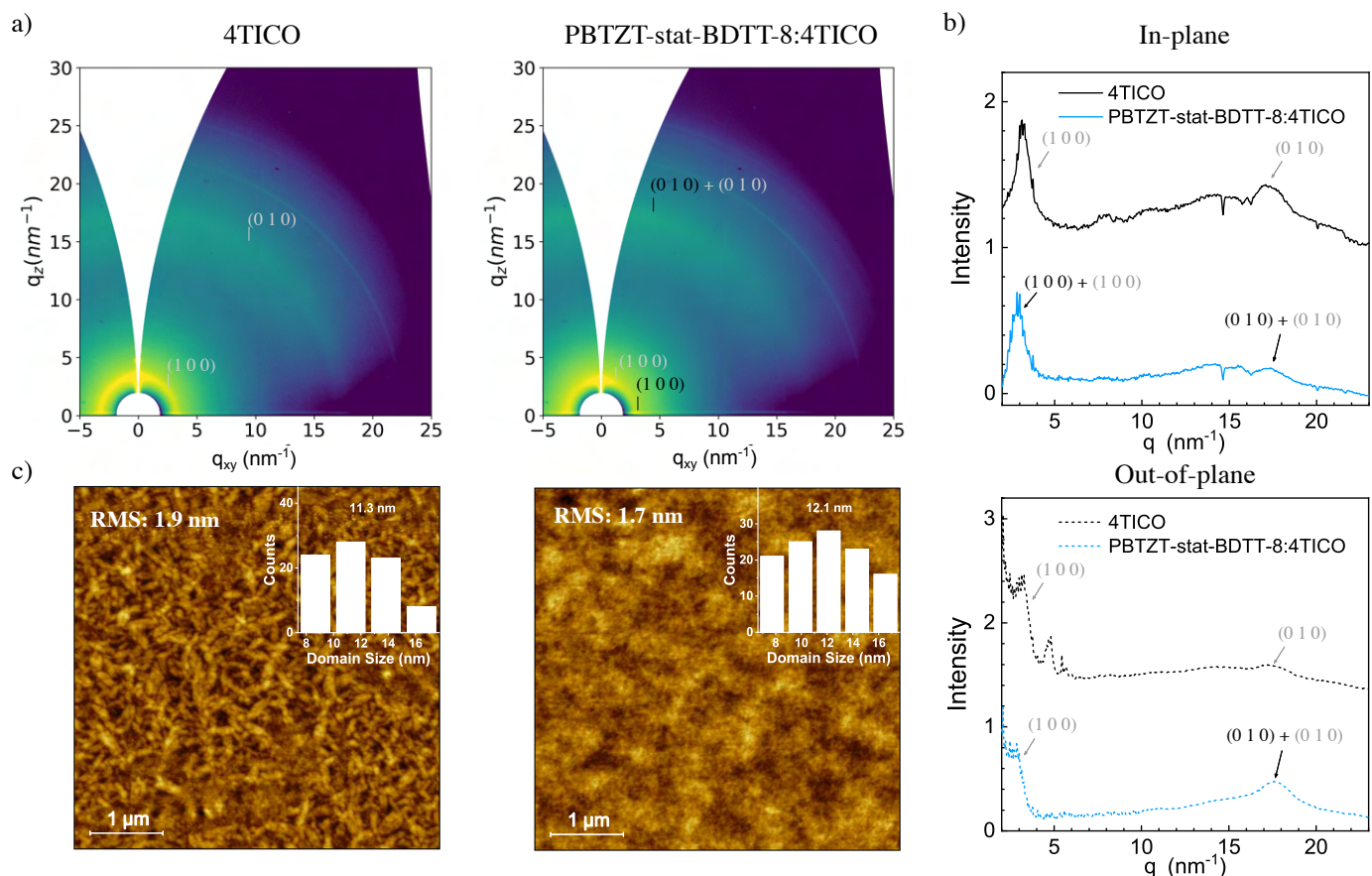

**Figure S13.** 4TICO and PBTZT-stat-BDTT-8:4TICO (a) GIWAXS patterns with in-plane and out-of-plane (b) integration profiles. 5x5 μm AFM images (c) of IDIC and PBTZT-stat-BDTT-8:IDIC films with domain size distribution and average value (inset).

**Table S8.** Crystallographic information of the main peaks observed by 2D-GIWAXS on 4TICO and PBTZT-stat-BDTT-8:4TICO films. AFM domain size and domain purity are also shown.

| Component | Peak | Orientation | q (nm$^{-1}$) | d (nm) | FWHM (nm$^{-1}$) | CCL (nm) | g | Domain Size (nm) | φ (%) |
|---|---|---|---|---|---|---|---|---|---|
| *NFA film* | | | | | | | | | |
| 4TICO | (1 0 0) | In-Plane | 3.20 | 1.96 | 0.89 | 6.3 | 21.0 | 11.3 | 56.2 |
| 4TICO | (0 1 0) | Out-of-plane | 17.55 | 0.36 | 3.09 | 1.8 | 16.7 | - | - |
| *Blend film* | | | | | | | | | |
| 4TICO | (1 0 0) | In-Plane | 3.12 | 2.01 | 0.98 | 5.8 | 22.4 | 12.1 | 47.7 |
| PBTZT-stat-BDTT-8 | (1 0 0) | In-plane | 2.77 | 2.27 | 0.64 | 8.8 | 19.2 | - | - |
| PBTZT-stat-BDTT-8:4TICO | (0 1 0) | Out-of-plane | 17.67 | 0.36 | 3.14 | 1.8 | 16.8 | - | - |

## ITIC

ITIC 2D-GIWAXS pattern (Figure S14a) is characterised by a weak crystallinity as it only presents one main feature with a lack of domain ordering along any specific direction. For this reason, it is not convenient to perform a side-by-side comparison with the single crystal structure.[5] However, a low angle (1 0 0) diffraction ring can still be distinguished (Figure S14a,b) and used for the evaluation of the NFA crystallinity in films (Table S9). As regards the PBTZT-stat-BDTT-8:ITIC blend, we could differentiate between the ITIC and PBTZT-stat-BDTT-8 lamellar reflections since the two are characterised by a different in-plane lattice spacing (Table S9 and Figure S14). We therefore conclude that the NFA is poorly crystalline in the blend given the high FWHM of the observed reflections with lack of directionality.

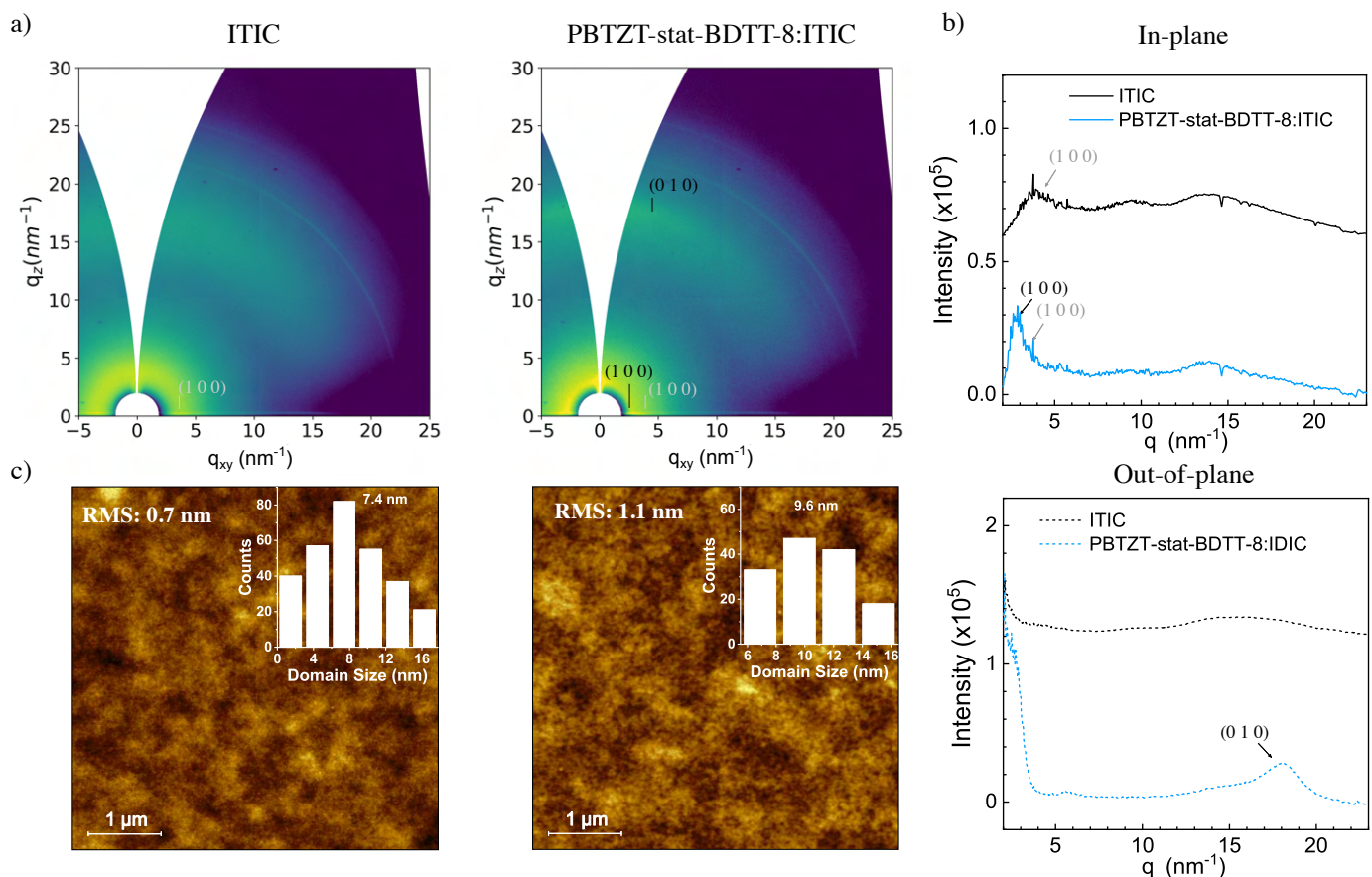

**Figure S14.** ITIC and PBTZT-stat-BDTT-8:ITIC (a) GIWAXS patterns with in-plane and out-of-plane (b) integration profiles. 5x5 µm AFM images (c) of ITIC and PBTZT-stat-BDTT-8:ITIC films with domain size distribution and average value (inset).

**Table S9.** Crystallographic information of the main peaks observed by 2D-GIWAXS on ITIC and PBTZT-stat-BDTT-8:ITIC films. AFM domain size and domain purity are also shown.

| Component | Peak | Orientation | q (nm$^{-1}$) | d (nm) | FWHM (nm$^{-1}$) | CCL (nm) | g | Domain Size (nm) | φ (%) |
|---|---|---|---|---|---|---|---|---|---|
| *NFA film* | | | | | | | | | |
| ITIC | (1 0 0) | In-Plane | 3.66 | 1.72 | 1.65 | 3.43 | 26.8 | 7.4 | 46.3 |
| *Blend film* | | | | | | | | | |
| ITIC | (1 0 0) | In-Plane | 3.57 | 1.76 | 0.99 | 5.7 | 21.0 | 9.6 | 59.4 |
| PBTZT-stat-BDTT-8 | (1 0 0) | In-plane | 2.76 | 2.28 | 0.79 | 7.1 | 21.3 | - | - |
| PBTZT-stat-BDTT-8:4TICO | (0 1 0) | Out-of-plane | 18.09 | 0.35 | 2.44 | 2.3 | 14.7 | - | - |

## 4. AFM Watershed Analysis

IDIC

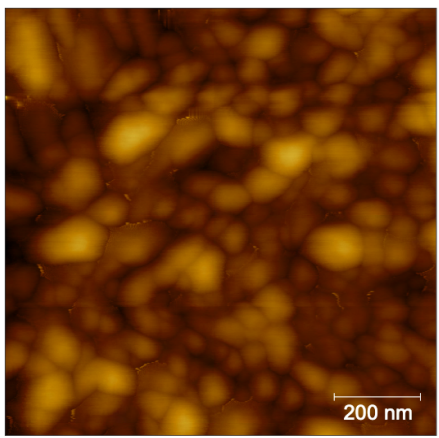 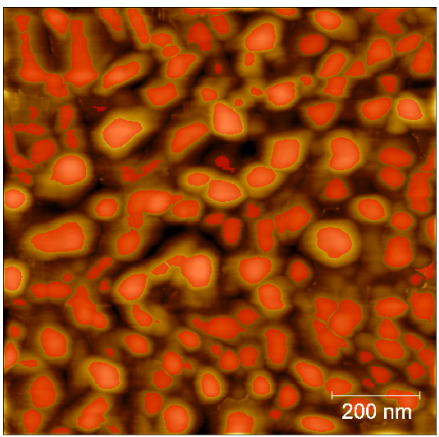

Avg. Domain Size: 52.7 nm

PBTZT-*stat*-BDTT-8:IDIC

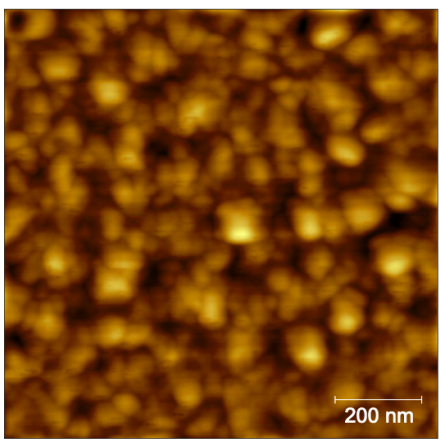 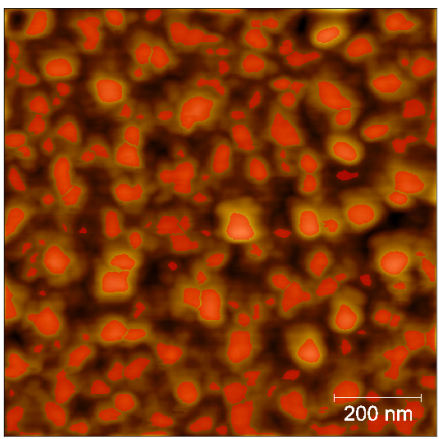

Avg. Domain Size: 41.1 nm

**Figure S15.** AFM images (1 μm x 1 μm) of IDIC and PBTZT-*stat*-BDTT-8:IDIC films with and without the mask used for the calculation of the average domain size by watershed algorithm. The input parameters and the result of the simulation are also shown.

*m*-ITIC

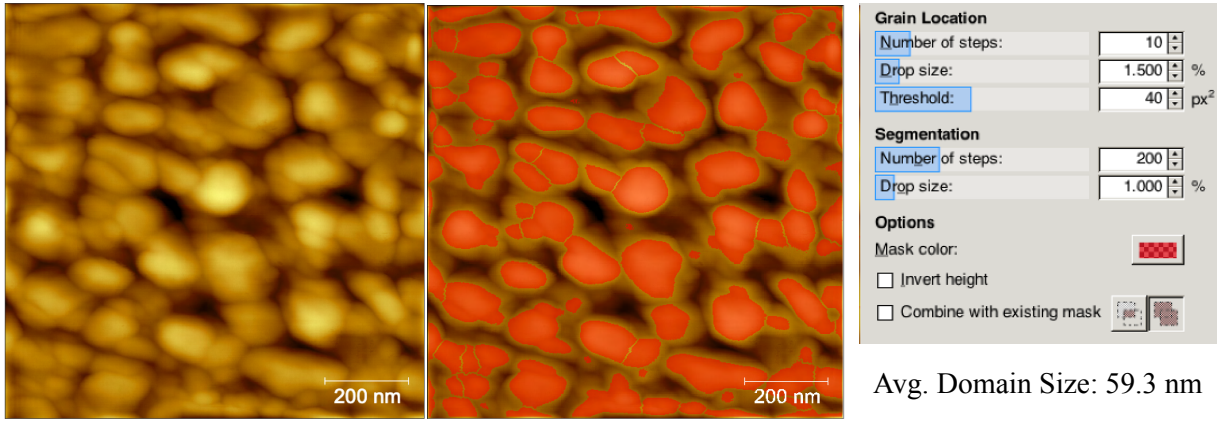

Avg. Domain Size: 59.3 nm

PBTZT-*stat*-BDTT-8:*m*-ITIC

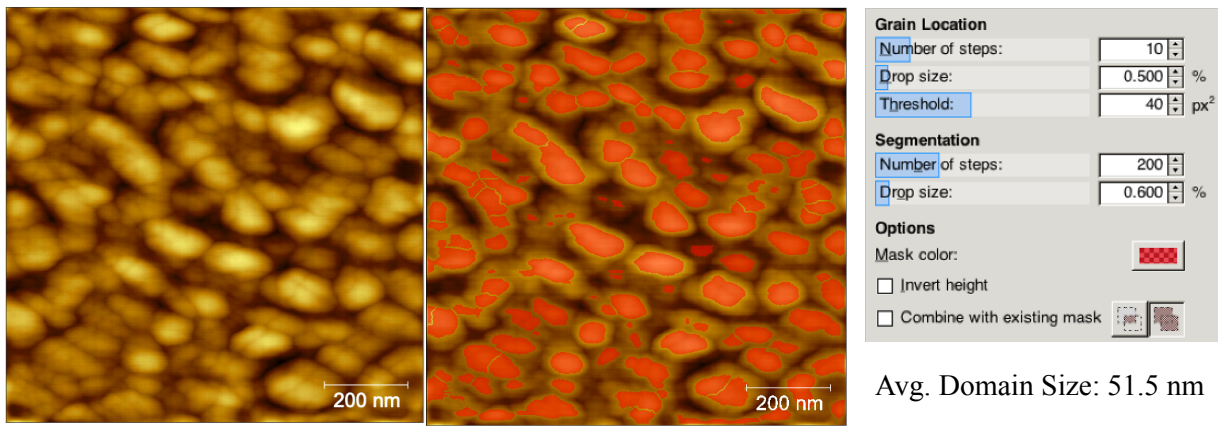

Avg. Domain Size: 51.5 nm

**Figure S16.** AFM images (1 μm x 1 μm) of *m*-ITIC and PBTZT-*stat*-BDTT-8:*m*-ITIC films with and without the mask used for the calculation of the average domain size by watershed algorithm. The input parameters and the result of the simulation are also shown.

*o*-IDTBR

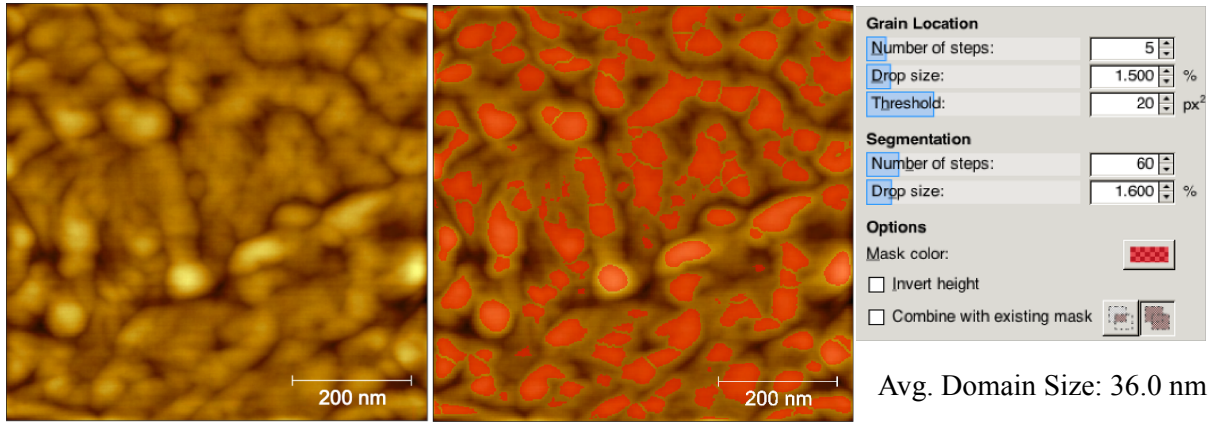

Avg. Domain Size: 36.0 nm

PBTZT-*stat*-BDTT-8:*o*-IDTBR

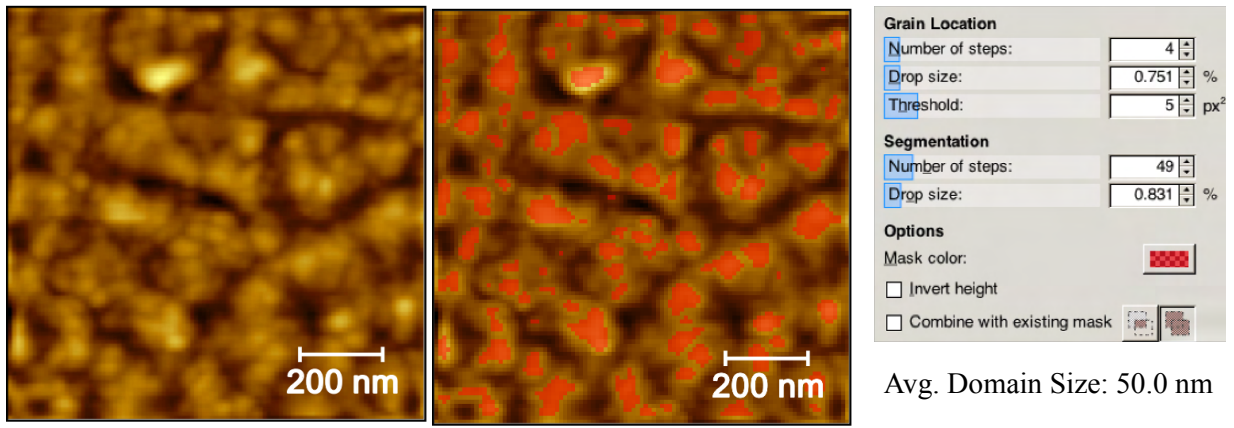

Avg. Domain Size: 50.0 nm

**Figure S17.** AFM images of *o*-IDTBR (0.6 μm x 0.6 μm) and PBTZT-*stat*-BDTT-8:*o*-IDTBR (1 μm x 1 μm) films with and without the mask used for the calculation of the average domain size by watershed algorithm. The input parameters and the result of the simulation are also shown.

*m*-4TICO

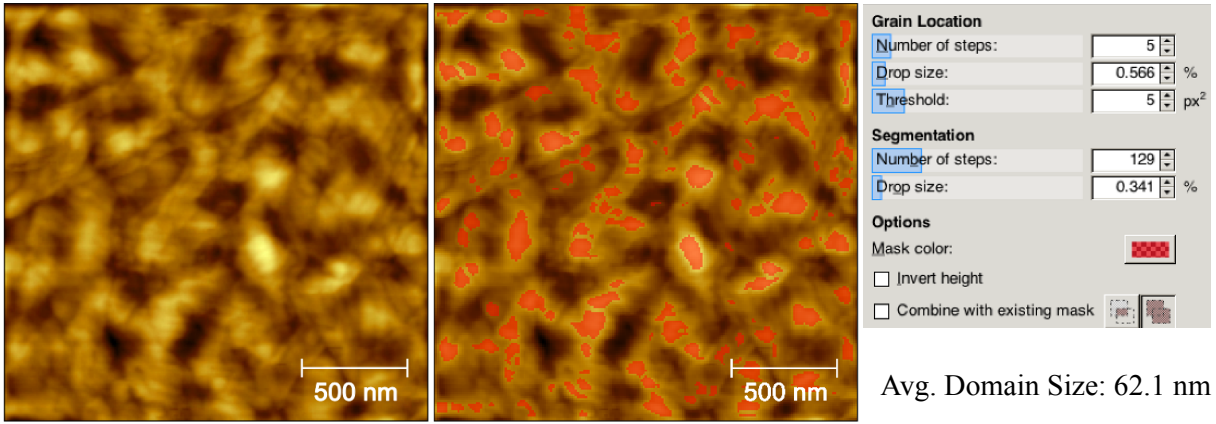

Avg. Domain Size: 62.1 nm

PBTZT-*stat*-BDTT-8:*m*-4TICO

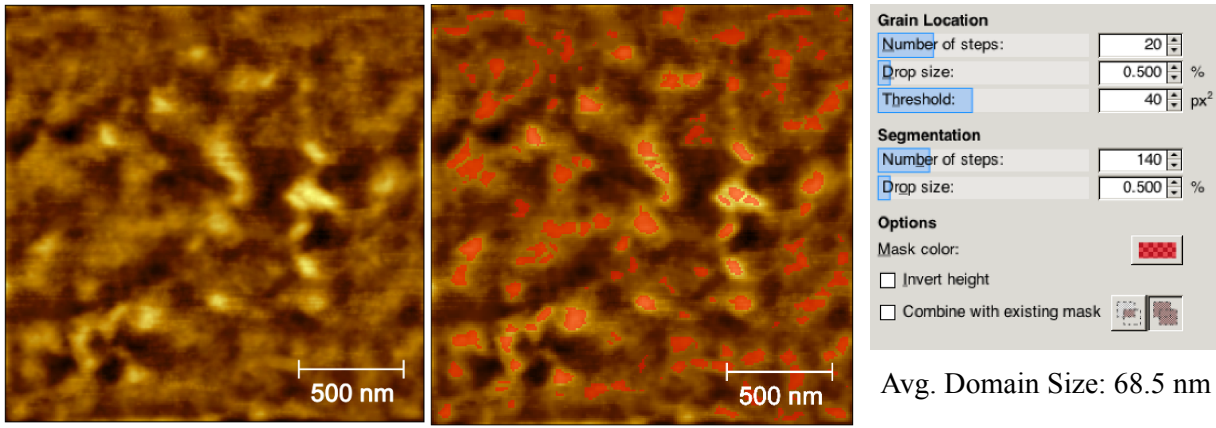

Avg. Domain Size: 68.5 nm

**Figure S18.** AFM images (2 μm x 2 μm) of *m*-4TICO and PBTZT-*stat*-BDTT-8:*m*-4TICO films with and without the mask used for the calculation of the average domain size by watershed algorithm. The input parameters and the result of the simulation are also shown.

ITIC

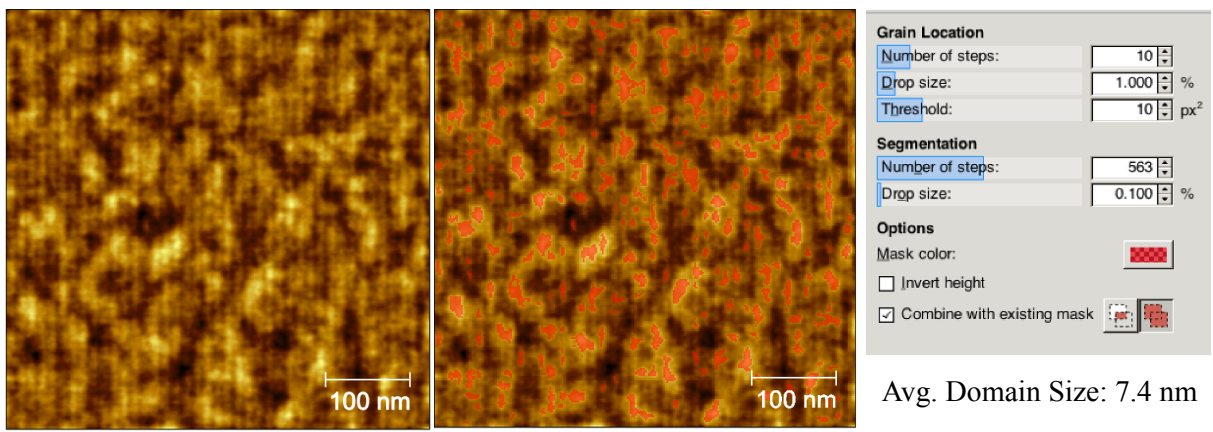

Avg. Domain Size: 7.4 nm

PBTZT-*stat*-BDTT-8:ITIC

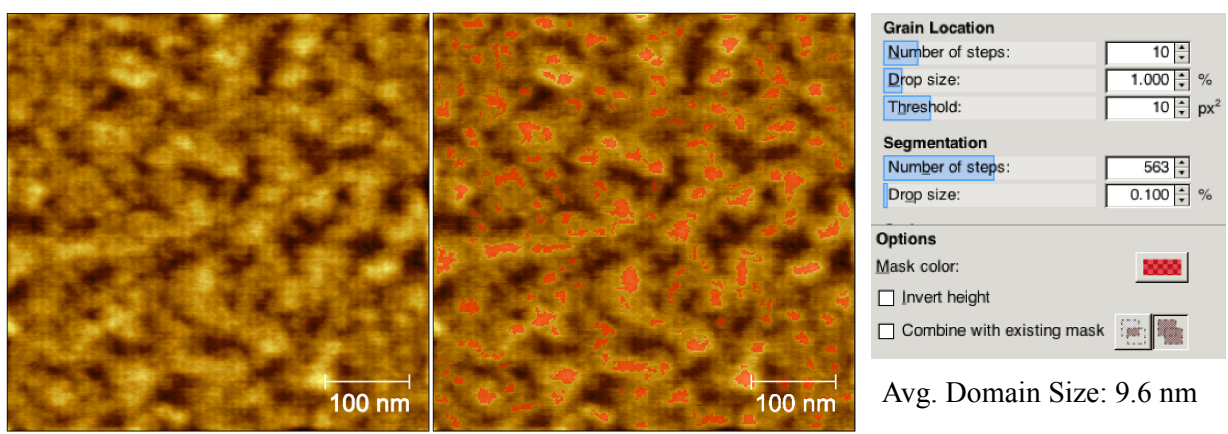

Avg. Domain Size: 9.6 nm

**Figure S19.** AFM images (0.5 μm x 0.5 μm) of ITIC and PBTZT-*stat*-BDTT-8:ITIC films with and without the mask used for the calculation of the average domain size by watershed algorithm. The input parameters and the result of the simulation are also shown.

## 4TICO

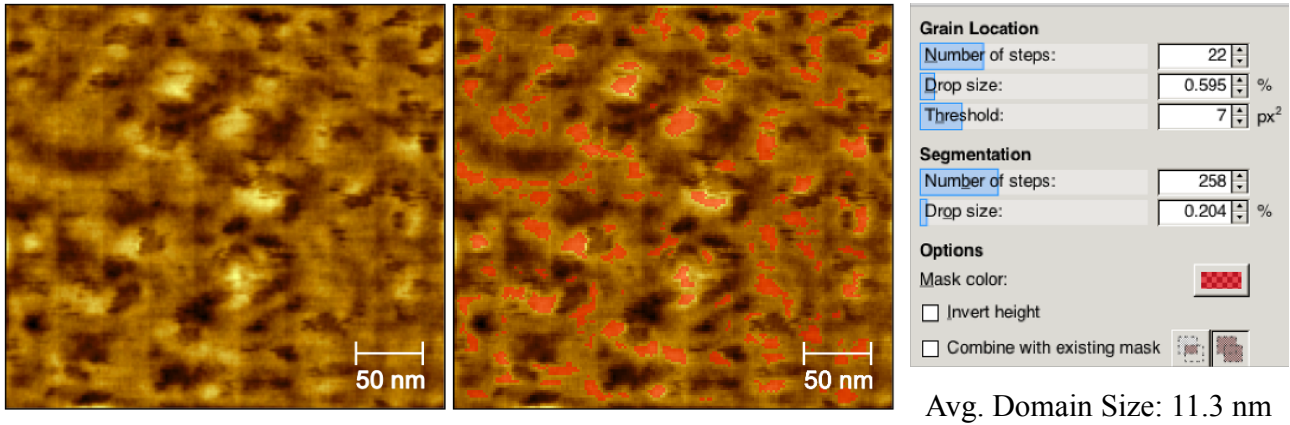

Avg. Domain Size: 11.3 nm

## PBTZT-*stat*-BDTT-8:4TICO

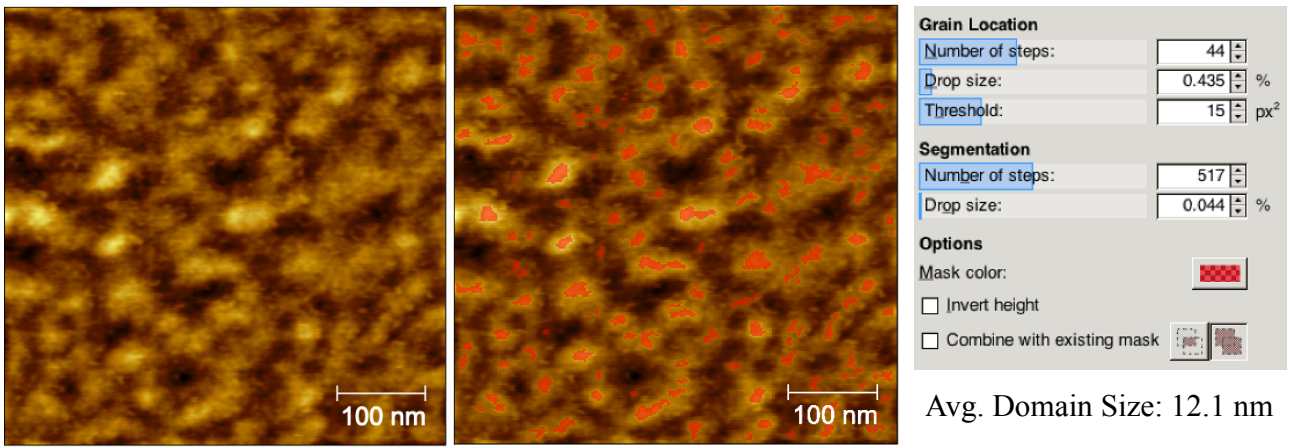

Avg. Domain Size: 12.1 nm

**Figure S20.** AFM images of 4TICO (0.3 μm x 0.3 μm) and PBTZT-*stat*-BDTT-8:4TICO (0.5 μm x 0.5 μm) films with and without the mask used for the calculation of the average domain size by watershed algorithm. The input parameters and the result of the simulation are also shown.

## PBTZT-*stat*-BDTT-8:4TIC

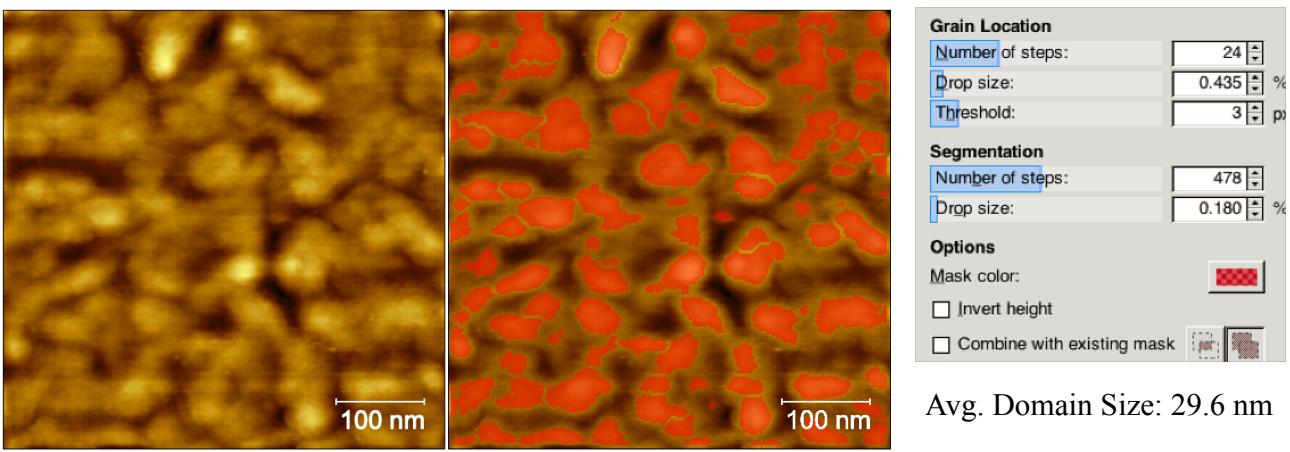

Avg. Domain Size: 29.6 nm

**Figure S21.** AFM images of PBTZT-*stat*-BDTT-8:4TIC (0.5 μm x 0.5 μm) films with and without the mask used for the calculation of the average domain size by watershed algorithm. The input parameters and the result of the simulation are also shown.

## 5. Electron Mobility by MIS-CELIV

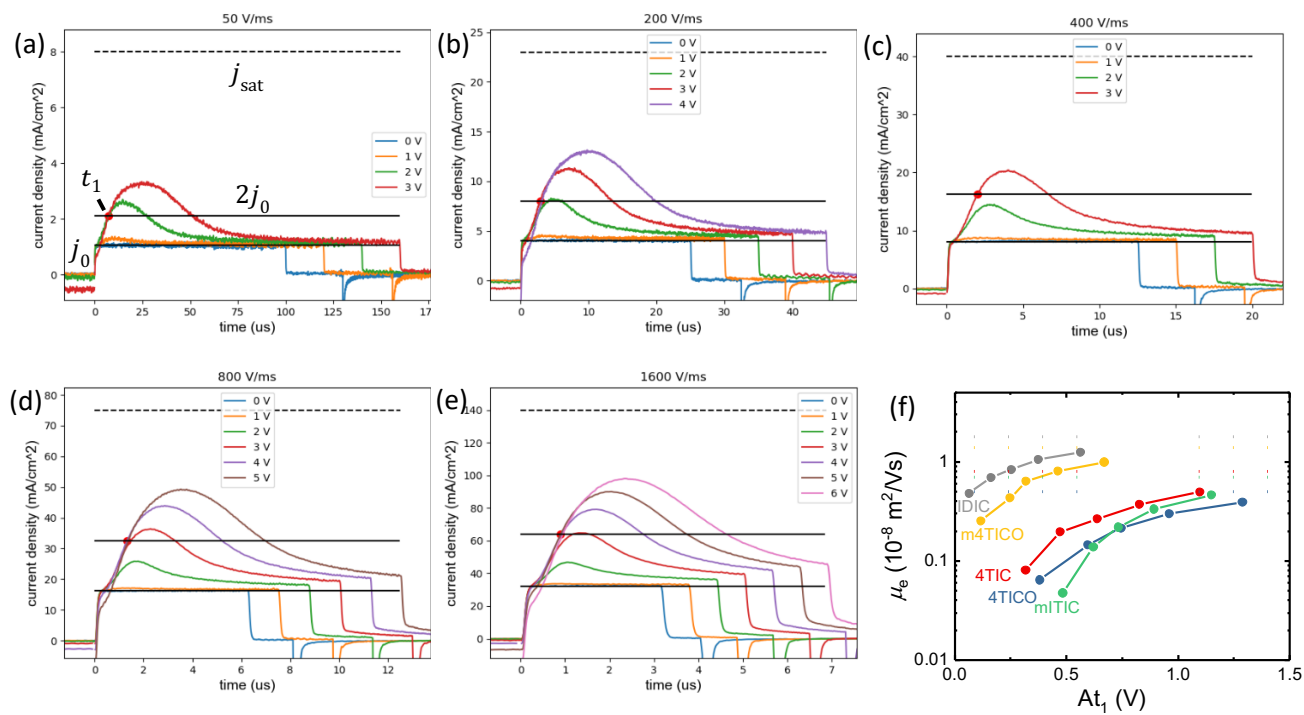

Figure S22 Typical MIS CELIV dataset taken for the 4TIC blend. The ramp rate *A* was varied between (a) 50 V/ms, (b) 200V/ms, (c) 400V/ms, (d) 800V/ms, and (e) 1600V/ms. (f) Extracted electron mobilities for the different blends in dependence of the ramp rate. Dashed lines indicated the estimated saturated mobility used in the main part. See text for further explanations.

Figure S15 shows representative example MIS-CELIV traces for PBTZT-stat-BDTT-8:4TIC. Different panels show measurements at different ramp rates and for each ramp rate the offset voltage is varied. The solid black lines indicate the displacement current density $j_0$ resulting from the geometric capacitance, and 2×$j_0$ used to determine time $t_1$ (red dot) in MIS-CELIV measurements. Finally, the saturated current density $j_{sat}$ is estimated by the top dashed line. Measurements at higher offset voltage suffered from increased current injection so that $j_{sat}$ had to be roughly estimated from the depicted traces. Note that the extracted mobility values change little with the exact value of $j_{sat}$. With $t_1$, the mobility is calculated for all ramp rates and blends and plotted in fig. S15 (bottom right panel). In the presence of some injection barrier, the measured apparent mobility depends on the product of ramp rate A and $t_1$. The measured mobility saturates and reaches its true value for large A×$t_1$.[6] The estimated saturated mobility is indicated with dashed lines in fig. S15 (bottom right panel).

With increasing annealing temperature, the blocking properties of MgF$_2$ diminished and for some samples leakage currents prevented data analysis. Hence, the m-4TICO blend at 100°C, rather than 120°C, and the unannealed m-ITIC were analysed.

However, we did not observe a notable change for other samples with annealing temperature. Hence, we can assume that the measured values are representative.

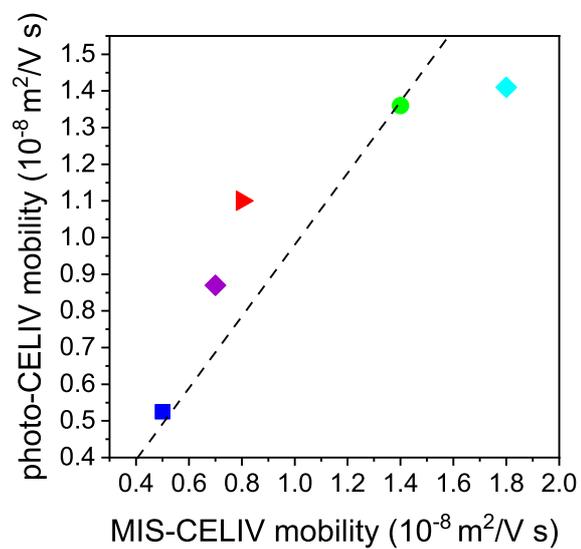

**Figure S23.** Photo-CELIV mobility compared to MIS-CELIV. A bisecting line is also plotted.

## 6. Active Layer Annealing Temperature and Dark J-V

**Table S10.** Thermal annealing protocol used for each active layer.

| Active Layer | Annealing Protocol |
|---|---|
| PBTZT-stat-BDTT-8:o-IDTBR | 5' at 100 °C |
| PBTZT-stat-BDTT-8:m-ITIC | 5' at 100 °C + 5' at 120 °C + 3' at 140 °C |
| PBTZT-stat-BDTT-8:ITIC | 5' at 100 °C + 5' at 120 °C + 3' at 140 °C |
| PBTZT-stat-BDTT-8:IDIC | 5' at 100 °C + 5' at 120 °C |
| PBTZT-stat-BDTT-8:4TICO | 5' at 100 °C + 5' at 120 °C + 3' at 140 °C |
| PBTZT-stat-BDTT-8:4TIC | 5' at 100 °C + 5' at 120 °C + 3' at 140 °C |
| PBTZT-stat-BDTT-8:m-4TICO | 5' at 100 °C + 5' at 120 °C |

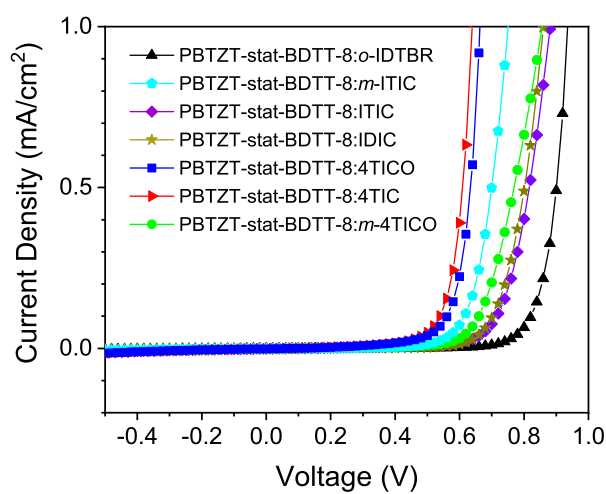

**Figure S24.** J-V characteristics measured in dark conditions.

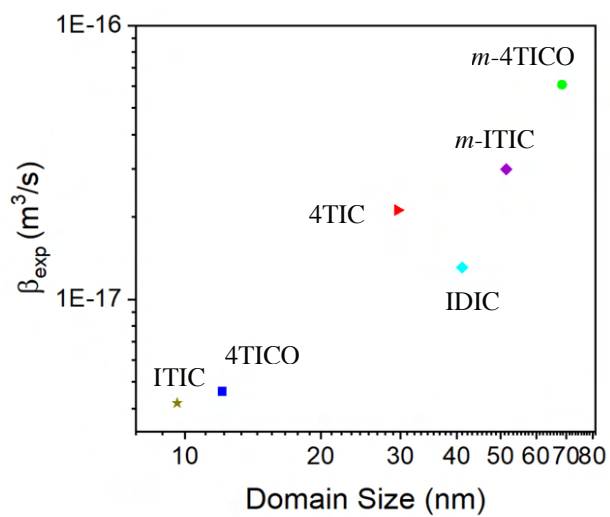

**Figure S25.** Plot showing the trend between the bimolecular recombination coefficient and the average domain size determined from AFM..